\documentclass[aps,pra,showkeys,twocolumn,superscriptaddress]{revtex4-2}
\usepackage[utf8]{inputenc}
\usepackage{amsmath,graphicx,array,float,amsthm}
\usepackage[most]{tcolorbox}
\usepackage{lipsum}
\usepackage[colorlinks={true}]{hyperref}
\hypersetup{colorlinks=true,linkcolor=red,citecolor=blue,urlcolor=blue}
\usepackage[FIGTOPCAP,raggedright,nooneline,bf,footnotesize]{subfigure}
\usepackage{indentfirst}
\usepackage{braket}
\usepackage{diagbox}
\usepackage{amssymb}
\usepackage{bbold}
\usepackage{colortbl}
\usepackage{multirow}
\usepackage{orcidlink}
\usepackage{pstricks}
\usepackage{amsmath}

\begin{document}
 \title{Quantum Characteristics Near Event Horizons}

\author{Asad Ali\orcidlink{0000-0001-9243-417X}}
\affiliation{Qatar Centre for Quantum Computing, College of Science and Engineering, Hamad Bin Khalifa University, Doha, Qatar}

\author{Saif Al-Kuwari\orcidlink{0000-0002-4402-7710}}
\affiliation{Qatar Centre for Quantum Computing, College of Science and Engineering, Hamad Bin Khalifa University, Doha, Qatar}

\author{Mehedad Ghominejad\orcidlink{0000-0002-0136-7838}}
\affiliation{Faculty of Physics, Semnan University, P.O. Box 35195-363, Semnan, Iran}

\author{M. T. Rahim\orcidlink{0000-0003-1529-928X}}
\affiliation{Qatar Centre for Quantum Computing, College of Science and Engineering, Hamad Bin Khalifa University, Doha, Qatar}

\author{Dong Wang\orcidlink{0000-0002-0545-6205}}
\affiliation{School of Physics and Optoelectronics Engineering, Anhui University, Hefei, 230601, People’s Republic of China}

\author{Saeed Haddadi\orcidlink{0000-0002-1596-0763}} \email{haddadi@semnan.ac.ir}
\affiliation{Faculty of Physics, Semnan University, P.O. Box 35195-363, Semnan, Iran}

\begin{abstract}
{ We investigate quantum characteristics around Schwarzschild black hole, exploring various quantum resources and their interplay in curved space-time. Our analysis reveals intriguing behaviors of quantum coherence, global and genuine multipartite entanglement, first-order coherence, and mutual information in different scenarios. Initially, we consider three particles shared among Alice, Bob, and Charlie in a Minkowski space far from the event horizon, where these particles are correlated via GHZ-type correlation. While Alice's particle remains in Minkowski space, Bob and Charlie accelerate towards the event horizon, experiencing black hole evaporation and generating antiparticles correlated via the Hawking effect. We employ the Kruskal basis formulation to derive a penta-partite pure state shared among particles inside and outside the event horizon. Investigating different scenarios among particles both inside and outside the event horizon, we observe how quantum resources evolve and distribute among consideration of different particles  with Hawking temperature and mode frequency. The trade-off relationship between first-order coherence and concurrence fill persists, indicating the intricate interplay between coherence and entanglement. Notably, the mutual information between external observers and particles inside the black hole becomes non-zero, deepening our understanding of quantum effects in curved space-time and shedding light on the quantum nature of the black hole. We believe that these findings will pave the way for future investigations into the fundamental quantum mechanical aspects of gravity under extreme environments.}
\end{abstract}


\maketitle

\section{Introduction}\label{sec1}
Black holes (BHs), stemming from Schwarzschild's solution to Einstein's general relativity, have captivated scientific inquiry since 1916 \cite{schwarzschild1916gravitationsfeld}. The groundbreaking release of the first BH image in 2019 by the Event Horizon Telescope marked a milestone \cite{akiyama2019first}. BHs, according to the no-hair theorem, appear to conceal information beyond their mass, charge, and angular momentum \cite{gurlebeck2015no}. However, the discovery of Hawking radiation by Stephen Hawking suggests a gradual evaporation of BH, raising questions about unitarity \cite{hawking1974black, denis2023entropy,almheiri2021entropy}. This phenomenon involves the creation of particle pairs near the event horizon, with one escaping and the other contributing to the BH's eventual disappearance \cite{PhysRevD.88.084010}. The event horizon of a BH is a boundary in space beyond which nothing, not even light, can escape the BH's gravitational pull. When an object or light crosses this boundary, it is inevitably pulled into the BH and its information is lost to external observers.

Even now, understanding the intricate interplay between quantum theory, general relativity, and the profound mysteries surrounding BH physics is a formidable challenge that has captivated the curiosity of physicists for decades \cite{DongWangADP2018,Huang2018Hawking,SHI2018649,li2022quantumness,Wu2022EPJC,Wu2022EPJC2,wu2023monogamy,wang2023nonperturbative,fujita2023holographic,Wu_2023,EPJC2023Fei}. This study embarks on a journey to investigate the qualitative migration and transformation of quantum resources in curved space-times, specifically focusing on quantifying genuine multipartite entanglement (GME), global entanglement, and quantum coherence among different particles and inside each particle in the context of Dirac fields interacting with a Schwarzschild BH.

Quantum information theory provides a unique lens for investigating foundational puzzles in relativistic quantum physics. Some basic concepts such as entanglement and coherence have proven instrumental in elucidating quantum effects in the perplexing environments near BHs \cite{xu2014hawking, martin2010unveiling, wang2010entanglement, wang2010projective,haddadi2024}. The enigma of the BH information paradox, revolving around the potential loss of information as matter crosses the event horizon, has been a focal point of inquiry. Hawking's initial calculations suggested information loss \cite{hawking1974black,denis2023entropy}, but in later work, Hawking proposed the escape of information through subtle quantum correlations in Hawking radiation \cite{hawking2005information}.

The study of entanglement between partitions around BHs has been extensive \cite{xu2014hawking, martin2010unveiling, wang2010entanglement, wang2010projective}, but global entanglement and GME can reveal richer multipartite correlations \cite{xie2021triangle}. Global concurrence (GC) \cite{meyer2002global,xie2021triangle, brennen2003observable} emerges as a quantifier that encapsulates both bipartite and multipartite entanglement contributions, providing a comprehensive measure of total entanglement between all parties involved. Additionally, examining first-order coherence (FOC) becomes crucial in capturing quantum superpositions within the local states of individual particles \cite{PhysRevLett.115.220501,svozilik2015revealing,ali2021properties, dong2022unification,PhysRevA.103.032407}. { From the understanding of the trade-off relation between FOC and concurrence-fill (CF), a genuine multipartite entanglement measure, as particles approach the BH, FOC may transition into CF through interactions as a trade-off
 \cite{dong2022unification}.} Hence, the GME measures are essential to specifically address irreducible multiparty inseparable correlations \cite{ guo2023complete, jin2023concurrence}. These measures quantify quantum correlations that are not reducible to any subset of particles and may offer a nuanced understanding of the intricate quantum fabric surrounding BHs.

This paper examines how Dirac field particles behave near a Schwarzschild BH, focusing on the average superposition of all the particles through FOC,  GME through concurrence fill (CF), and global entanglement through GC in curved space-time. { In order to see if quantum correlations other than entanglement exist, we also quantify the $l_1$-norm of quantum coherence (QC).} Additionally, the study looks into how the Hawking temperature and mode frequency of Dirac particles near the BH affect these quantum behaviors and aims to explain how these quantum resources manifest trade-offs in curved space-time.
The meticulous tracking of these quantum resources not only deepens our understanding of information dynamics but also offers valuable insights into the intersection of quantum theory and general relativity.

\section{Preliminaries}\label{sec2}
{ This section provides the definitions and some necessary properties of quantum coherence (captured by $l_1$-norm of QC and FOC), global entanglement (captured by CF and GC),  and mutual information.}
{
\subsection{$l_1$-norm of quantum coherence}
QC arises from the fundamental superposition principle in quantum mechanics. A rigorous framework to quantify coherence as a resource has been developed, known as the resource theory of QC \cite{baumgratz2014quantifying,HU20181}. This theory identifies the set of incoherent states $\mathcal{I}$ which are diagonal in a reference basis $\{|i\rangle\}$:
\begin{equation}
\delta \in \mathcal{I} \iff \delta = \sum_i \delta_i |i\rangle\langle i|.
\label{eq1}
\end{equation}
The free operations are the incoherent operations that map incoherent states to incoherent states. Revealing and quantifying QC is essential to enable quantum correlations and information processing. Hence, Baumgratz et al. \cite{baumgratz2014quantifying} proposed the $l_1$-norm of QC as a quantifier of coherence:
\begin{equation}
C(\rho) = \sum_{i\neq j} |\langle i|\rho|j\rangle| = \sum_{i,j} |\rho_{ij}| - \sum_i |\rho_{ii}|.
\end{equation}
\label{eq2}

\subsection{First-order coherence}
Pauli matrices $\bold{\sigma}=(\sigma_1,\sigma_2,\sigma_3)$ together with the identity matrix $I$, provide a complete set of operator bases in Liouville space to express any general two-qubit density matrix $\rho$ in the following parameterized form:
\begin{equation}
\begin{aligned}
\rho = & \frac{1}{4} \Big(I \otimes I + \bold{r'}_x \cdot \bold{\sigma} \otimes I \\
& + I \otimes \bold{r'}_y \cdot \bold{\sigma} + \sum_{m,n=1}^{3} c_{mn}~\sigma_m \otimes \sigma_n\Big).
\end{aligned}
\label{eq3}
\end{equation}
Here, $c_{mn}=\textmd{tr}(\rho\sigma_m\otimes\sigma_n)$ is the matrix element of matrix $C\in\Re^{m\times n}$, and $\bold{r'}_x$ and $\bold{r'}_y$ are the Bloch vectors corresponding to each qubit. The unitary equivalent form  of $\rho$ under local unitary transformation $U \otimes V$ can be written as
\begin{align}
\begin{split}
\rho = & U \otimes V \rho_{xy} U^\dagger \otimes V^\dagger
= \frac{1}{4} \Big(I \otimes I + \bold{r}_x \cdot \bold{\sigma} \otimes I \\
& + I \otimes \bold{r}_y \cdot \bold{\sigma} + \sum_{i=1}^{3} c_{i}\sigma_i \otimes \sigma_i\Big),
\end{split}
\label{eq4}
\end{align}
 where $\bold{r}_x=\textmd{tr}[U(\bold{r'}_x.\bold{\sigma})U^\dagger]\bold{\sigma}$ and $\bold{r}_y=\textmd{tr}[V(\bold{r'}_y.\bold{\sigma})V^\dagger]\bold{\sigma}$ are the corresponding local unitary equivalent Bloch vectors, and $c_i$ are the eigenvalues of $3\times3$ $C^\dagger C$ matrix.

 Using reduced states $\rho_x=\frac{1}{2}(I+\bold{r}_x\cdot\bold{\sigma})$ and $\rho_y=\frac{1}{2}(I+\bold{r}_y\cdot\bold{\sigma})$, one can define the FOC of individual reduced states $D(\rho_x)=|\bold{r}_x|=\sqrt{2 \textmd{tr}(\rho_{x}^{2})-1}$ and $D(\rho_y)=|\bold{r}_y|=\sqrt{2 \textmd{tr}(\rho_{y}^{2})-1}$. Therefore, the FOC of $\rho$, i.e. $D(\rho)$, would be written as mean square averages of $D(\rho_x)$ and $D(\rho_y)$, given by \cite{PhysRevLett.115.220501,dong2022unification, svozilik2015revealing, ali2021properties,PhysRevA.103.032407}
\begin{align}
D(\rho)= \sqrt{\frac{|\bold{r}_x|^2+|\bold{r}_y|^2}{2}}=\sqrt{\textmd{tr}({\rho}^2_x)+\textmd{tr}({\rho}^2_y)-1}.
\label{eq5}
\end{align}

Based on the fact that $\textmd{tr}(\rho^2_i) \geq \frac{1}{2}$, where $i \in {x, y}$, we can find that $\textmd{tr}(\rho^2_x) + \textmd{tr}(\rho^2_y) \geq 1$ which assures that  $0\leq D(\rho)\leq1$.

Considering all subsystems as independent entities, the total FOC of the tripartite state can be generalized as the root mean square average of all FOC of the individual subsystems \cite{dong2022unification}, namely
\begin{equation}
D(\rho_{xyz})=\sqrt{\frac{D^{2}(\rho_{x})+D^{2}(\rho_{y})+D^{2}(\rho_{z})}{3}},
\label{eq6}
\end{equation}
with $0\leq D(\rho_{xyz})\leq1$.

Note that FOC in a tripartite system is defined as the root mean squared average of the individual local hidden coherence, given by $D(\rho_i)=\sqrt{2 \textmd{tr}(\rho_{i}^{2})-1}$. The local coherence of any subsystem indicates that in a maximally mixed state (represented by $\frac{1}{2} I$), coherence is absent, with purity expressed as $\textmd{tr}(\rho_i^2)=1/2$, signifying maximal mixedness. Conversely, for a maximally pure quantum state, coherence reaches unity, corresponding to $\textmd{tr}(\rho_i^2)=1$.

Recent research has unveiled a functional complementarity trade-off relationship of FOC with various metrics such as quantum non-locality \cite{svozilik2015revealing}, quantum steering \cite{dong2022unification}, separability (separable uncertainty) \cite{tessier2005complementarity}, and concurrence \cite{fan2019universal,PhysRevA.107.052403}. It is hypothesized that FOC can provide insights into the distribution of quantum coherence among subsystems within a multipartite system, elucidating quantum correlations in terms of entanglement, steering, nonlocality, and other phenomena.
}
\subsection{Concurrence fill and global concurrence}
GME involves non-separable quantum correlations among three or more particles that cannot be simplified into pairwise entanglements \cite{xie2021triangle, guo2023complete, jin2023concurrence, guo2023complete}. This complex entanglement exceeds bipartite forms between just two particles. When an $N$-particle state is indivisible into separate parts, it exhibits genuine $N$-partite entanglement. To qualify as a measure, any GME measure must meet certain criteria, including \cite{guo2022genuine, guo2023complete, jin2023concurrence, guo2023complete}:
\begin{itemize}
\item If a multipartite quantum state $\rho$ belongs to the set of bi-separable states $S_{\textmd{bi-sep}}$, then the GME measure, denoted by $F(\rho)$, should be zero, i.e. $F(\rho)=0$. Conversely, if the state $\rho$ is closed under the set of GME carrying states $S_{\textmd{GME}}$ (non-bi-separable states), then $F(\rho)$ is anticipated to be greater than zero, i.e. $F(\rho)>0$. Specifically, the normalized GME measure should satisfy $F(\rho)=1$ for a maximally genuine multipartite entangled state. Therefore, in a general context, we can express this relationship as follows
\begin{equation}
0\leq F(\rho)\leq1.
\label{eq7}
\end{equation}

\item When considering an ensemble of quantum states ($p_{i},\rho_{i}$) obtained through local operation and classical communication (LOCC) applied to the initial state $\rho$, the GME measure is expected to adhere to the following monotonicity condition
\begin{equation}
F(\rho) \geq \sum_{i} p_{i} F(\rho_{i}).
\label{eq7}
\end{equation}
This inequality signifies that under LOCC operations, the GME measure is monotonic.

\item For any arbitary unitary operator $U$, the GME measure must preserve unitarity, namely
\begin{equation}
F(U\rho U^{\dagger})=F(\rho).
\label{eq8}
\end{equation}
\end{itemize}

Recently, progress has been made in determining the proper order for genuine tripartite entanglement by introducing the concept of CF.

The interconnection among three bipartite entanglements is interrelated, where one system is involved with the other two. These entanglements are not mutually independent but adhere to a specific relationship \cite{zhu2015generalized}, as shown
\begin{equation}
C^{2}_{x(yz)}\leq C^{2}_{y(zx)}+C^{2}_{z(xy)},
\label{eq9}
\end{equation}
where
\begin{equation}
C_{i(j k)}=2\sqrt{\det(\rho_{i})},
\label{eq10}
\end{equation}
with $0\leq C_{i(j k)}\leq1$ where $i,~j,~k \in \{x,~y,~z\}~\forall i\neq j\neq k$.

This inequality captures the squares of three bipartite concurrences, resembling the lengths of sides in a triangle called the ``concurrence triangle". The CF is subsequently defined { for pure states} as the square root of the area enclosed by this so-called concurrence triangle as \cite{xie2021triangle}
\begin{align}
\begin{split}
    F(\left|\psi\right\rangle) &= \bigg\{\frac{16}{3}Q(\left|\psi\right\rangle) \\
    &\quad\times \left[Q(\left|\psi\right\rangle)-C^2_{x(yz)}(\left|\psi\right\rangle)\right] \\
    &\quad\times \left[Q(\left|\psi\right\rangle)-C^2_{y(zx)}(\left|\psi\right\rangle)\right] \\
    &\quad\times \left[Q(\left|\psi\right\rangle)-C^2_{z(xy)}(\left|\psi\right\rangle)\right]\bigg\}^{1/4},
\end{split}
\label{eq11}
\end{align}
where
\begin{equation}
Q(\left|\psi\right\rangle)=\frac{1}{2}\left[
C^{2}_{x(yz)}(\left|\psi\right\rangle)+ C^{2}_{y(zx)}(\left|\psi\right\rangle)+C^{2}_{z(xy)}(\left|\psi\right\rangle)\right],
\label{eq12}
\end{equation}
is the half-perimeter of the concurrence triangle from Heron's formula, also known as GC \cite{meyer2002global,xie2021triangle, brennen2003observable}, while the pre-factor 16/3 ensures the normalization condition that is $0\leq F(\rho_{xyz}) \leq1$.

{ Notably, CF \eqref{eq11} and GC \eqref{eq12} can be generalized to the case of mixed states through the convex roof construction, given by \cite{xie2021triangle}
\begin{equation}
F(\rho_{xyz})=\min _{\left\{p_i, \psi_i\right\}} \sum_i p_i F(\left|\psi_i\rangle\right),
\label{eq13}
\end{equation}
and
\begin{equation}
Q(\rho_{xyz})=\min _{\left\{p_i, \psi_i\right\}} \sum_i p_i Q(\left|\psi_i\rangle\right),
\label{eq14}
\end{equation}
in which the minimum is taken over all possible decompositions $\rho_{xyz}=\sum_i p_i\left|\psi_i\right\rangle\left\langle\psi_i\right|$.}

{
It is worth mentioning that recently a mathematical trade-off relation between FOC and CF has been established \cite{dong2022unification}, given by
\begin{equation}
    D^2(\rho_{xyz})+F(\rho_{xyz})\leq1.
    \label{eq15}
\end{equation}
}

\begin{figure}[t]
	\begin{center}     \includegraphics[width=0.43\textwidth]{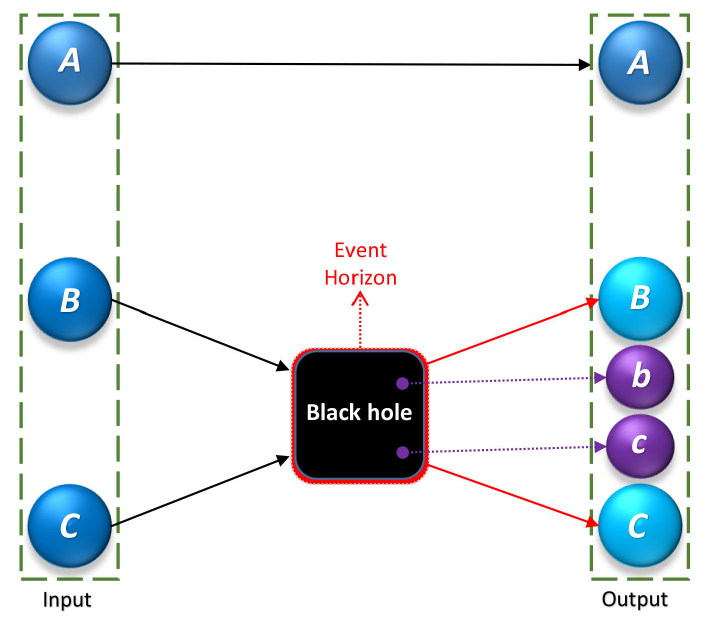}	
	\end{center}
	\caption{Schematic diagram of our physical model with Alice's particle--$A$ in a flat region, and Bob's particle--$B$ and Charlie's particle--$C$ near the event horizon of a Schwarzschild BH. The dashed lines show the entanglement between particles. Input state is provided in \eqref{eq23} and output state is given in \eqref{5qubits}.}
	\label{fig1}
\end{figure}

{ \subsection{Mutual Information}
Understanding the correlations between subsystems is fundamental in quantum information theory. Quantum mutual information, denoted by $I_{\rho_{XY}}$, serves as a vital tool for quantifying this correlation. Consider a bipartite system composed of subsystems $\rho_{X}$ and $\rho_{Y}$. Their state space can be described by the tensor product of their individual Hilbert spaces, denoted by $H_X \otimes H_Y$.

The quantum mutual information, $I_{\rho_{XY}}$, captures the amount of information shared between $\rho_X$ and $\rho_Y$. It is mathematically expressed as
\begin{equation}
I_{\rho_{XY}} = S(\rho_X) + S(\rho_Y) - S(\rho_{XY}),
\label{eq16}
\end{equation}
where
$\rho_{XY}$ represents the density matrix of the entire system residing in $H_X \otimes H_Y$.
$\rho_X = \textmd{tr}_Y(\rho_{XY})$ and $\rho_Y = \textmd{tr}_X(\rho_{XY})$ are the reduced density matrices of $\rho_{XY}$ for subsystems $X$ and $Y$, respectively. Here,
$S(\rho) = - \textmd{tr}(\rho \log_2 \rho)$ denotes the von Neumann entropy of a density matrix $\rho$, which quantifies the uncertainty or mixedness of the quantum state.

Notice that the von Neumann entropy of the reduced states, $S(\rho_X)$ and $S(\rho_Y)$, represents the information inherent in each subsystem individually. The entropy of the whole system, $S(\rho_{XY})$, captures the combined information of both $X$ and $Y$. In general, the quantum mutual information, $I_{\rho_{XY}}$, essentially reflects the information gained about one subsystem (say, $X$) by knowing the state of the other ($Y$).}

\section{Quantum Treatment of Dirac Field}\label{sec3}
The metric in the background of a Schwarzschild space-time can be specified as
\begin{align}
ds^{2} &= -\left(1-\frac{2M}{r}\right)dt^{2} + \left(1-\frac{2M}{r}\right)^{-1}dr^{2} \nonumber \\
&\quad+ r^{2}(d\theta^{2} + \sin^{2}\theta d\varphi^{2}),
\label{eq17}
\end{align}
where $M$ represents the mass of BH. For convenience, we work in the natural units, for which $G=c=\hbar=k_B=1$. When considering a general background space-time, the Dirac equation would be expressed as \cite{xu2014hawking}
{\small\begin{align}
& -\frac{\gamma_{0}}{\sqrt{1-\frac{2M}{r}}}\frac{\partial\Phi}{\partial t}
 +\gamma_{1}\sqrt{1-\frac{2M}{r}}\left[\frac{\partial}{\partial r}+\frac{1}{r}+\frac{M}{2r(r-2M)}\right]\Phi \nonumber \\
& +\frac{\gamma_{2}}{r}\left(\frac{\partial}{\partial\theta}+\frac{\cot\theta}{2}\right)\Phi +\frac{\gamma_{3}}{r\sin\theta}\frac{\partial\Phi}{\partial\varphi} = 0,
\label{eq18}
\end{align}}
where $\gamma_{i}$ $(i=0,1,2,3)$ represent Dirac gamma matrices.
A set of positive-frequency outgoing solutions can be obtained by solving the Dirac equation, as expressed in Eq. \eqref{eq18} near the BH's event horizon. These solutions are relevant for describing the event horizon's interior and exterior regions as
\begin{equation}
\Phi_{k,in}^{+}=\phi(r) e^{i\omega\tau}
\label{eq19}
\end{equation}
and
\begin{equation}
\Phi_{k,out}^{+}=\phi(r) e^{-i\omega\tau}.
\label{eq20}
\end{equation}

In the given context, $\phi(r)$ indicates a four-component Dirac spinor, $\omega$ denotes a monochromatic frequency, $k$ is the wave vector, and $\tau$ is defined as $t-r^*$ with $r^*$ being the tortoise coordinate given by $r^*=r+2M\ln\frac{r-2M}{2M}$. Note that the modes identified as $\Phi_{k, \text{in}}^{+}$ and $\Phi_{k, \text{out}}^{+}$ are commonly known as Schwarzschild modes.

Following Damour and Ruffini's suggestion \cite{damour1976black}, we can extend the given equation analytically, establishing a solid basis for positive energy modes. This extension enables the derivation of Bogoliubov transformations \cite{bogoljubov1958new, barnett2002methods, xu2014hawking} related to the creation and annihilation operators in both Schwarzschild and Kruskal coordinates. By quantizing Dirac fields in the Schwarzschild and Kruskal modes and appropriately normalizing the state vector, one can articulate the formulations for the Kruskal vacuum and excited states with mode $k$ as \cite {xu2014hawking}
\begin{equation}
\left|0\right\rangle _{k}=S_{-}\left|0\right\rangle _{o}\left|0\right\rangle _{i}+S_{+}\left|1\right\rangle _{o}\left|1\right\rangle _{i}
\label{eq21}
\end{equation}
and
\begin{equation}
\left|1\right\rangle _{k}=\left|1\right\rangle _{o}\left|0\right\rangle _{i},
\label{eq22}
\end{equation}
where $S_{\pm}=(e^{\pm \omega/T_H}+1)^{-1/2}$ with the Hawking temperature as $T_H=1/8 \pi M$. Furthermore, $\left|f\right\rangle_{o}$ and $\left|f\right\rangle _{i}$ with $f=0,1$ are the Fock states for the particle pair outside the region with momentum $+k$ and inside the region with momentum $-k$ of the BH, respectively.

\begin{figure*}[t]
  \centering  \includegraphics[width=0.38\textwidth]{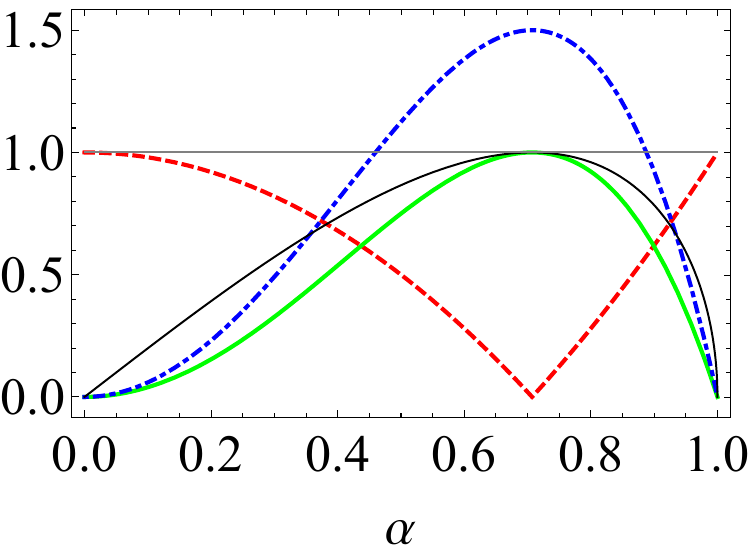}\put(-25,125){(a)}\qquad  \includegraphics[width=0.38\textwidth]{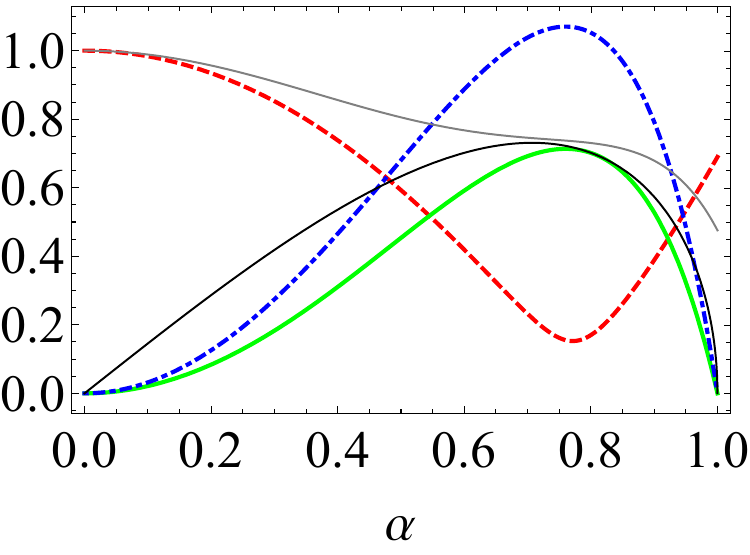} \put(-25,125){(b)}\qquad  \includegraphics[width=0.38\textwidth]{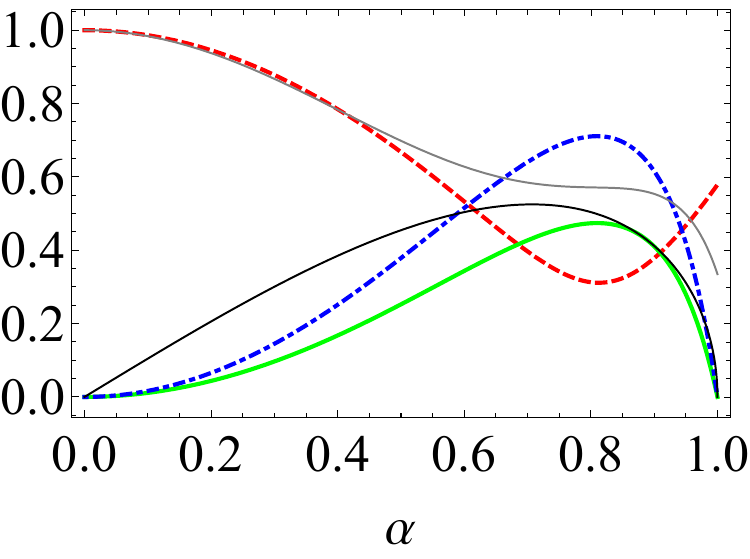}\put(-25,125){(c)}\qquad  \includegraphics[width=0.38\textwidth]{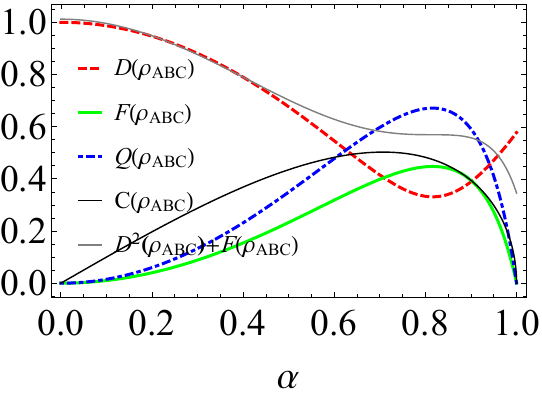}\put(-25,125){(d)}
\caption{FOC [$D(\rho_{ABC})$] (dashed-red), CF [$F(\rho_{ABC})$] (solid-green), GC [$Q(\rho_{ABC})$] (dot-dashed blue), QC [$C(\rho_{ABC})$] (solid-black) and $D^2(\rho_{ABC})+F(\rho_{ABC})$ (thin-solid gray)  as a function of $\alpha$ for $\omega=1$ at  $T_H=0.01$ (a), $T_H=1$ (b), $T_H=10$ (c),  and $T_H=100$ (d).}
\label{fig2}
\end{figure*}

\begin{figure*}[t]
\begin{center}		
\includegraphics[width=0.38\textwidth]{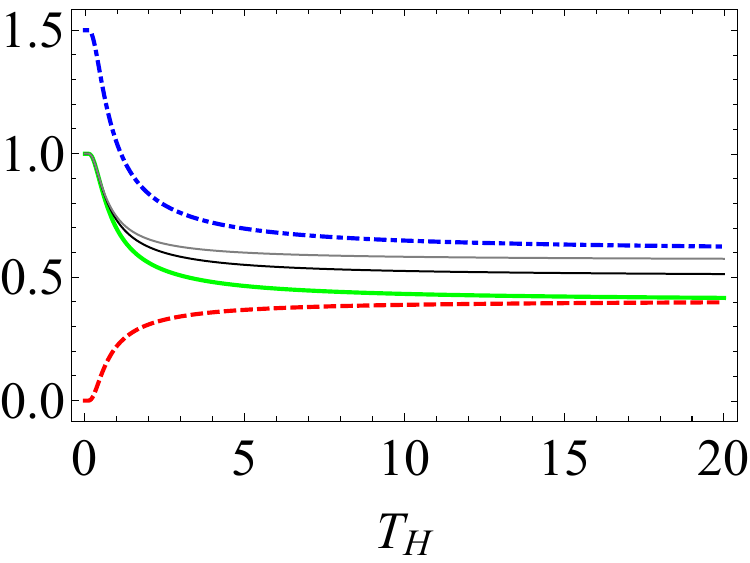}
\put(-160,125){(a)}\qquad	
\includegraphics[width=0.38\textwidth]{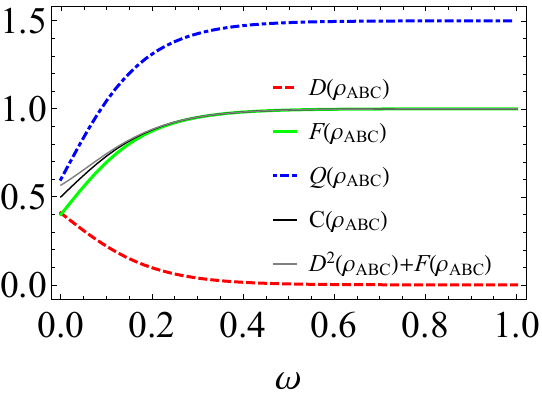}
\put(-160,125){(b)}
\end{center}
\caption{FOC [$D(\rho_{ABC})$] (dashed-red), CF [$F(\rho_{ABC})$] (solid-green), GC [$Q(\rho_{ABC})$] (dot-dashed blue), QC [$C(\rho_{ABC})$] (Solid-black) and $D^2(\rho_{ABC})+F(\rho_{ABC})$ (thin-solid gray) as functions of $T_H$ and $\omega$ with $\alpha=1/\sqrt{2}$. (a) $\omega=1$   and (b)  $T_H=0.1$.}
\label{fig3}
\end{figure*}

\begin{figure*}[!t]
\centering  \includegraphics[width=0.38\textwidth]{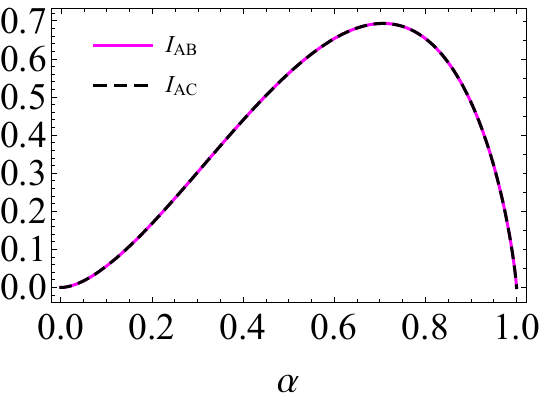}\put(-40,50){(a)}\qquad  \includegraphics[width=0.38\textwidth]{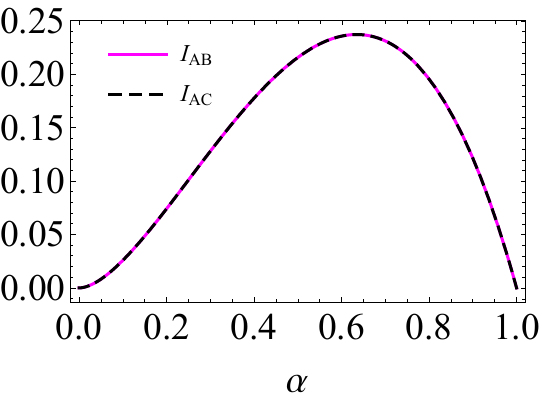} \put(-40,50){(b)}\qquad  \includegraphics[width=0.38\textwidth]{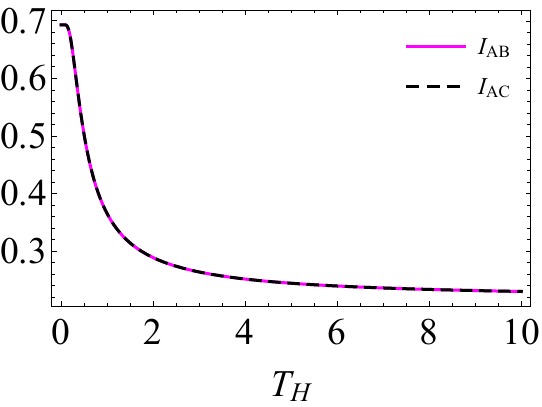}\put(-40,50){(c)}\qquad  \includegraphics[width=0.38\textwidth]{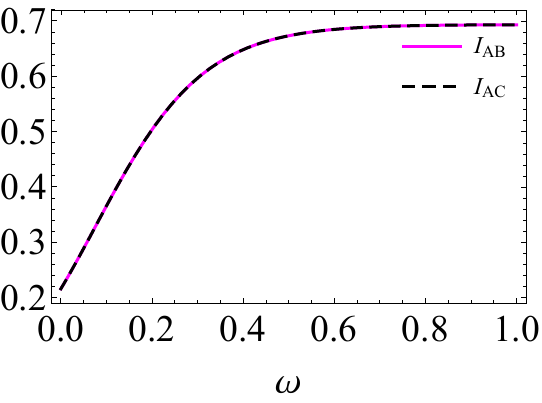}\put(-40,50){(d)}
\caption{Mutual information $I_{AB}$ and $I_{AC}$ based on state \eqref{eq24}  with (a) $\omega=1$; $T_H=0.01$,  (b) $\omega=1$; $T_H=10$, (c) $\omega=1$; $\alpha=1/\sqrt{2}$,  and (d) $T_H=0.1$; $\alpha=1/\sqrt{2}$.}
\label{fig4}
\end{figure*}

\section{Results and discussion}\label{sec4}
Entangled tripartite states, such as the GHZ-like state, are valuable quantum resources exhibiting GME.
In \cite{xie2021triangle}, it is shown that the GHZ state is the strongest GME-carrying state as well as being a maximally global entanglement-carrying state. Therefore, we establish and examine the GHZ-type state shared between three observers Alice, Bob, and Charlie in a flat Minkowski space-time outside the event horizon of a Schwarzschild BH. Let us assume that Alice's qubit is in $\left|f\right\rangle _{A}$, while Bob's and Charlie's qubits are in $\left|f\right\rangle _{B}$ and $\left|f\right\rangle _{C}$ respectively where $f$ can take two values, i.e. $0$ or $1$. The initial tripartite state shared between them can be written as
\begin{equation}
\left|\psi\right\rangle _{ABC}=\alpha\left|0_{A}0_{B}0_{C}\right\rangle +\sqrt{1-\alpha^{2}}\left|1_{A}1_{B}1_{C}\right\rangle,
\label{eq23}
\end{equation}
where $\alpha$ is the state parameter with $0\leq\alpha\leq1$.
{ As mentioned before, the initial state shared among Alice, Bob, and Charlie, denoted $\left|\psi\right\rangle_{ABC}$, is a GHZ-type pure state, which transitions into a GHZ state when $\alpha=\frac{1}{\sqrt{2}}$. This GHZ state is identified as a genuine maximally entangled three-qubit pure state. }

We now consider a scenario where Alice remains in the flat asymptotic region outside the event horizon, but Bob and Charlie fall freely toward the event horizon. Their respective antiparticles, Anti-Bob and Anti-Charlie, are located inside the event horizon with modes $\left|f\right\rangle_{b}$ and $\left|f\right\rangle_{c}$. Using the Kruskal basis shown in  Eqs. \eqref{eq21} and \eqref{eq22} for Bob and Charlie while treating Alice on a Minkowski basis, we can reformulate the complete penta-partite quantum state as (see Fig.  \ref{fig1})
\begin{align}\label{5qubits}
    \left|\psi\right\rangle _{AbBcC} =&\Theta_+ \left|0_{A}1_{b}1_{B}1_{c}1_{C}\right\rangle+ \Theta_- \left|0_{A}0_{b}0_{B}0_{c}0_{C}\right\rangle\nonumber \\
&+ \Gamma\big\{\left|0_{A}0_{b}0_{B}1_{c}1_{C}\right\rangle+\left|0_{A}1_{b}1_{B}0_{c}0_{C}\right\rangle\big\}\nonumber \\
     &+ \Upsilon \left|1_{A}0_{b}1_{B}0_{c}1_{C}\right\rangle,
\end{align}
where $\Theta_{\pm} = \alpha S_{\pm}^2,$ $\Gamma =\alpha/2 \sqrt{\cosh ^2\left(\omega/2 T_H\right)}$, and $ \Upsilon = \sqrt{1-\alpha ^2}$.

In a broader context, this quantum state embodies a pure five-partite entanglement, encompassing separate subsystems. Qubit $A$ undergoes observation by Alice, whereas qubits $B$ and $C$ are scrutinized by Bob and Charlie, respectively, positioned beyond the event horizon of the BH. Furthermore, qubits $b$ and $c$ fall under the observation of anti-Bob and anti-Charlie inside the event horizon.
Owing to the causal disconnection between the interior and exterior domains of the BH, Alice, Bob, and Charlie are devoid of access to the modes within the event horizon. Therefore, we classify the modes $B$ and $C$ outside the event horizon as the ``accessible modes" and the modes $b$ and $c$ inside the event horizon as the ``inaccessible modes." The process involves taking the trace over the inaccessible and accessible modes on $\left|\psi\right\rangle _{AbBcC}$ given in Eq. \eqref{5qubits}, resulting in the tripartite reduced density operators for different configurations.

Now, we explore all the possibilities of sharing the tripartite coherence and entanglement between different parties, both accessible and partially accessible. We consider three different scenarios: in the first scenario, three particles are accessible, in the next scenario two particles are accessible, and in the last scenario, only one particle is accessible.

\subsection{Alice--Bob--Charlie}\label{subsubsec:A}

Let us consider the accessible mode case comprised of Alice, Bob, and Charlie, whose density operator $\rho_{ABC}$ can be evaluated by taking the partial trace over anti-Bob and anti-Charlie modes given in Eq. \eqref{5qubits}, namely $\rho_{ABC}=\textmd{tr}_{bc}(\left|\psi\right\rangle_{AbBcC}\langle\psi|)$.  This yields
\begin{align}
    \rho_{ABC}=&\Theta_+^2 \left|0 1 1\right\rangle \left\langle 0 1 1\right| +\Theta_-^2 \left|0 0 0\right\rangle \left\langle 0 0 0\right|+\Upsilon^2 \left|1 1 1\right\rangle \left\langle 1 1 1\right|
     \nonumber \\
    &+\Upsilon \Theta_- \{ \left|0 0 0\right\rangle \left\langle 1 1 1\right|
    + \left|1 1 1\right\rangle \left\langle 0 0 0\right|\} \nonumber \\
    &+\Gamma ^2 \{\left|0 0 1\right\rangle \left\langle 0 0 1\right|
    + \left|0 1 0\right\rangle \left\langle 0 1 0\right|\}.
    \label{eq24}
\end{align}

{
Figure \ref{fig2}(a-d) shows the variation of $l_1$-norm of QC, GC, CF, and FOC concerning $\alpha$, with various fixed values of Hawking temperature at $\omega=1$. At $T_{H}=0.01$, depicted in Fig. \ref{fig2}(a), where nearly no Hawking radiation is present, we have an entangled state. Notably, a perfect trade-off between FOC and CF is evident, satisfying the upper bound relationship $D^2(\left|\psi\right\rangle _{ABC})+F(\left|\psi\right\rangle _{ABC})=1$. The peaks of QC, CF, and GC occur at $\alpha=1/\sqrt{2}$. Conversely, when $T_{H}=1$, $T_{H}=10$, and $T_{H}=100$, as depicted in Fig. \ref{fig2}(b), (c), and (d) respectively, the peaks diminish compared to Fig. \ref{fig2}(a), and the minimum value of FOC rises from zero. The trade-off persists without reaching the upper bound, i.e. $D^2(\rho_{ABC})+F(\rho_{ABC})<1$. This trend continues with even lower peak values of QC, CF, and GC, accompanied by increasing minimum values of FOC. Moreover, with an escalation in Hawking temperature, the maximum values of QC, GC, and CF shift towards $\alpha\approx0.8$, while the trade-off between FOC and CF remains satisfied. Notably, there is no significant decline in the metric values while transitioning from $T_H=10$ [Fig. \ref{fig2}(c)] to $T_H=100$ [Fig. \ref{fig2}(d)].

Figure \ref{fig3}(a) showcases the modulation of QC, GC, CF, and FOC concerning $T_H$ at $\omega=1$ and $\alpha=\frac{1}{\sqrt{2}}$. It is evident that with an increase in the Hawking temperature, QC, GC, and CF decrease from their maximum values to certain minimum values, thereafter stabilizing and persisting without reaching zero. In contrast, FOC begins from zero, reaching a certain value, and then saturating. Note that the trade-off relation between FOC and CF remains intact. This observation suggests that although Hawking temperatures degrade entanglement for a completely accessible scenario, they fail to entirely annihilate it, even for infinitely large Hawking temperatures.

Figure \ref{fig3}(b) illustrates the behaviors of QC, GC, CF, and FOC versus $\omega$ at $T_H=0.1$ and $\alpha=1/\sqrt{2}$. It is observed that with an increase in mode frequency, QC, GC, and CF ascend to maximum saturated values, maintaining their peak levels. Conversely, FOC decreases from its maximum value to zero at higher mode frequencies, all the while adhering to the trade-off between CF and FOC.

After examining the variations of QC, GC, CF, and FOC among Alice, Bob, and Charlie in a completely accessible scenario, we now investigate the mutual information shared between Alice and Bob, represented by $I_{AB}=S(\rho_A)+S(\rho_B)-S(\rho_{AB})$, and between Alice and Charlie, denoted by $I_{AC}=S(\rho_A)+S(\rho_C)-S(\rho_{AC})$, across different parameters. Figs. \ref{fig4}(a) and \ref{fig4}(b) illustrate $I_{AB}$ and $I_{AC}$ as a function of $\alpha$ at $T_{H}=0.01$ and $T_{H}=10$, respectively, for $\omega=1$. Remarkably, we observe the equivalence of $I_{AB}$ and $I_{AC}$  when $\rho_{AB}=\rho_{AC}$ (see appendix). Both $I_{AB}$ and $I_{AC}$ exhibit a similar trend to other metrics discussed in Fig. \ref{fig2} at $\alpha=1/\sqrt{2}$ with $T_{H}=0.01$. Fig. \ref{fig4}(c) reveals that both $I_{AB}$ and $I_{AC}$ decrease with increasing Hawking temperature and rise with mode frequency as depicted in Fig. \ref{fig4}(d), ultimately saturating for larger values of both parameters $T_H$ and $\omega$. Notably, as we have seen, increasing the temperature and frequency of the Hawking mode did not completely destroy quantum coherence and quantum entanglement.
}

\begin{figure*}[!t]
\centering  \includegraphics[width=0.38\textwidth]{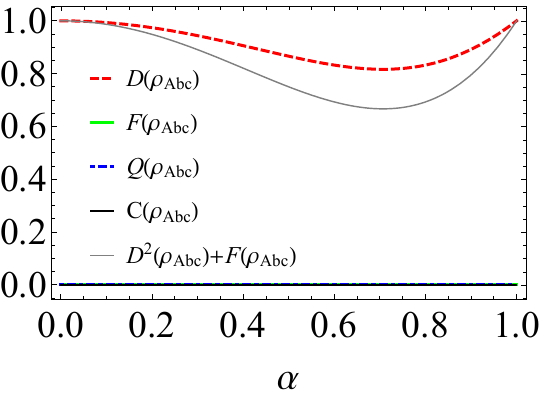}\put(-30,50){(a)}\qquad  \includegraphics[width=0.38\textwidth]{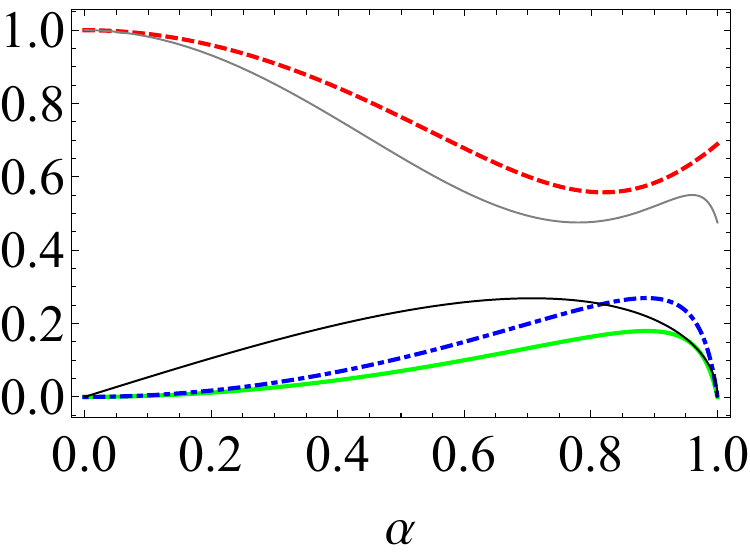} \put(-30,45){(b)}\qquad  \includegraphics[width=0.38\textwidth]{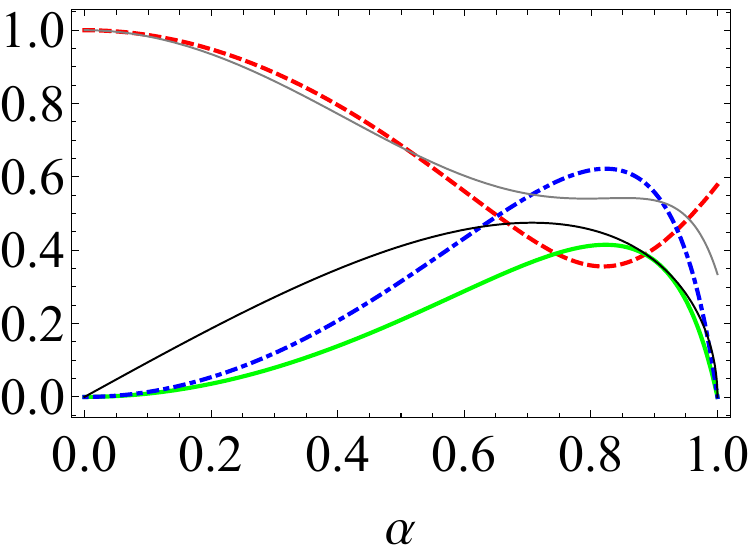}\put(-30,45){(c)}\qquad  \includegraphics[width=0.38\textwidth]{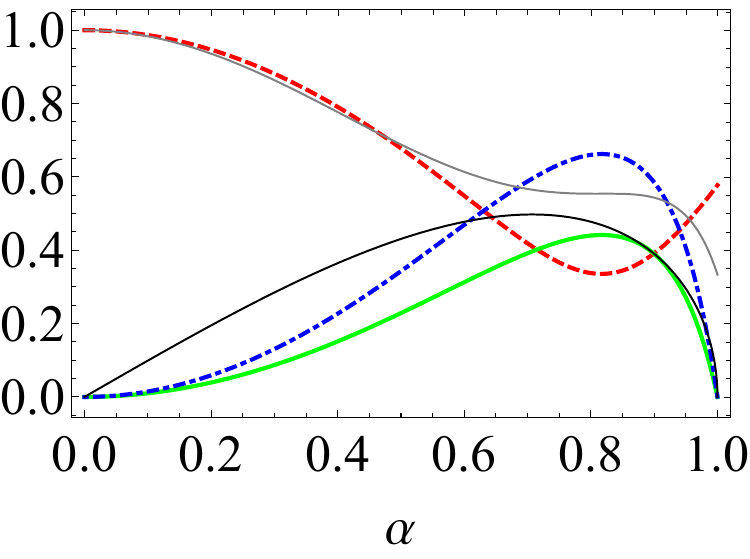}\put(-30,45){(d)}
\caption{FOC [$D(\rho_{Abc})$] (dashed-red), CF [$F(\rho_{Abc})$] (solid-green), GC [$Q(\rho_{Abc})$] (dot-dashed blue), QC [$C(\rho_{Abc})$] (solid-black) and $D^2(\rho_{Abc})+F(\rho_{Abc})$ (thin-solid gray)  as a function of $\alpha$ for $\omega=1$ at  $T_H=0.01$ (a), $T_H=1$ (b), $T_H=10$ (c),  and $T_H=100$ (d).}
\label{fig5}
\end{figure*}

\begin{figure*}[t]
\begin{center}		\includegraphics[width=0.38\textwidth]{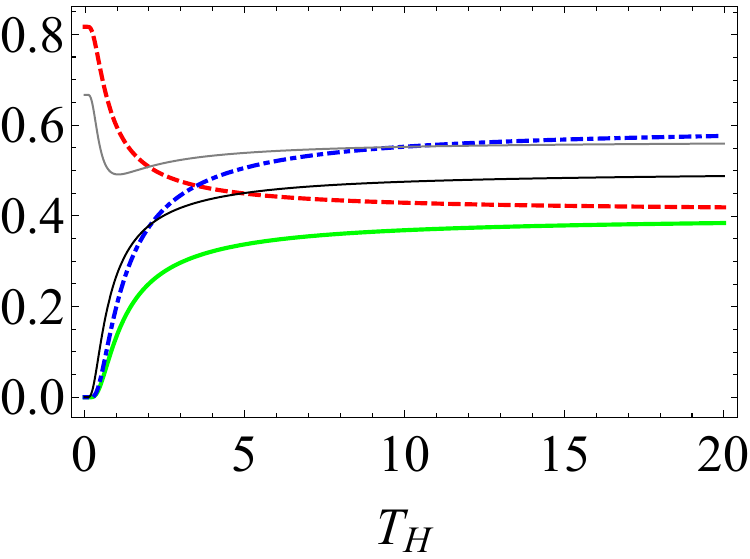}
		\put(-30,125){(a)}\qquad			\includegraphics[width=0.38\textwidth]{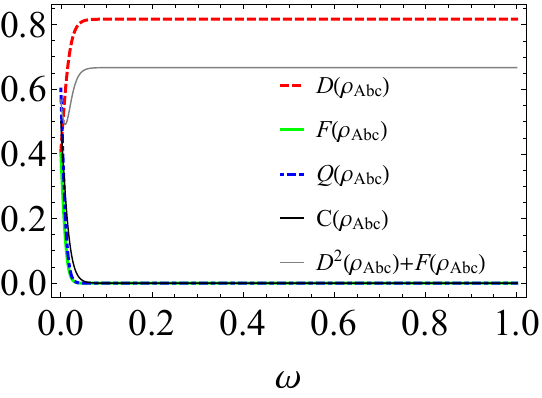}
		\put(-30,125){(b)}
	\end{center}
  \caption{FOC [$D(\rho_{Abc})$] (dashed-red), CF [$F(\rho_{Abc})$] (solid-green), GC [$Q(\rho_{Abc})$] (dot-dashed blue), QC [$C(\rho_{Abc})$] (Solid-black) and $D^2(\rho_{Abc})+F(\rho_{Abc})$ (thin-solid gray) as functions of $T_H$ and $\omega$ with $\alpha=1/\sqrt{2}$. (a) $\omega=1$   and (b)  $T_H=0.1$.}
  \label{fig6}
\end{figure*}

\begin{figure*}[!t]
\centering  \includegraphics[width=0.38\textwidth]{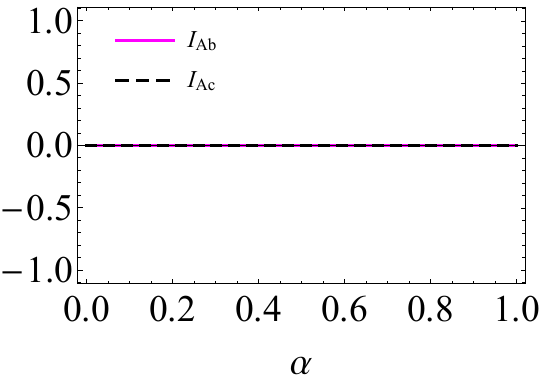}\put(-30,45){(a)}\qquad  \includegraphics[width=0.38\textwidth]{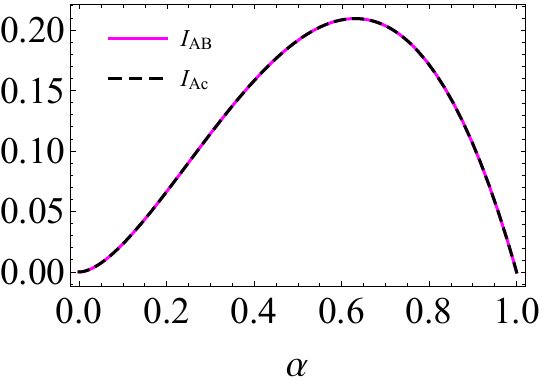} \put(-30,45){(b)}\qquad  \includegraphics[width=0.38\textwidth]{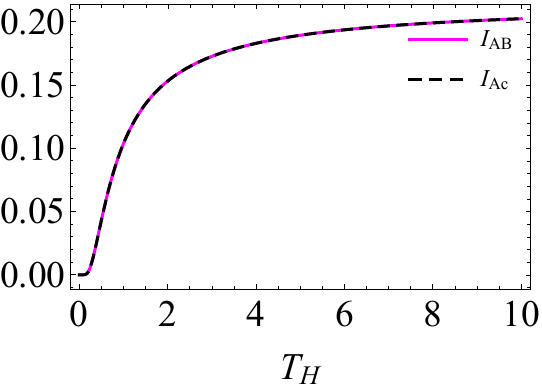}\put(-30,45){(c)}\qquad  \includegraphics[width=0.38\textwidth]{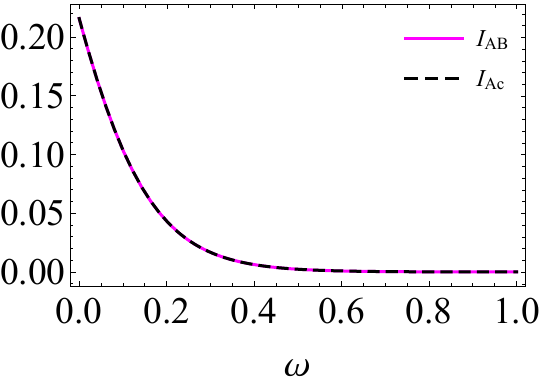}\put(-30,45){(d)}
\caption{Mutual information $I_{Ab}$ and $I_{Ac}$ based on state \eqref{eq25}  with (a) $\omega=1$; $T_H=0.01$,  (b) $\omega=1$; $T_H=10$, (c) $\omega=1$; $\alpha=1/\sqrt{2}$,  and (d) $T_H=0.1$; $\alpha=1/\sqrt{2}$.}
\label{fig7}
\end{figure*}

\subsection{Alice--anti-Bob--anti-Charlie}

Defining the interior of a BH is inherently challenging to explore practically, as an external observer encounters perturbative limitations, preventing the reception of signals from beyond the event horizon. However, we know that, in the unitary quantum mechanics framework, information preservation is obligatory.

Considering a scenario where two particles, referred to as anti-Bob and anti-Charlie, exist within the BH while Alice remains outside, though the physical exploration inside the BH is physically impractical, the complete state of our penta-partite system is known and expressed in Eq. \eqref{5qubits} as a pure state, maintaining unitarity. Consequently, the application of a partial tracing operation on the modes of Bob and Charlie within this penta-partite state yields $\rho_{Abc}$, given by
\begin{equation}
\begin{aligned}
    \rho_{Abc}=& \Theta_-^2 \left|000\right\rangle  \left\langle 000\right|+\Theta_+^2 \left|011\right\rangle  \left\langle 011\right| +  \Upsilon^2 \left|100\right\rangle  \left\langle 100\right| \\
    &+ \Theta_+ \Upsilon \{\left|011\right\rangle  \left\langle 100\right| +
 \left|100\right\rangle  \left\langle 011\right|\} \\
    &+ \Gamma^2 \{\left|001\right\rangle  \left\langle 001\right| +  \left|010\right\rangle  \left\langle 010\right|\}. \label{eq25}
\end{aligned}
\end{equation}

{
In Fig. \ref{fig5}, the modulation of QC, GC, CF, and FOC concerning $\alpha$ is illustrated, with various fixed values of Hawking temperature at $\omega=1$. In Fig. \ref{fig5}(a), for negligible Hawking temperature ($T_{H}=0.01$), the absence of Dirac particle-antiparticle pair production on the event horizon is observed. Consequently, in contrast to Fig. \ref{fig2}(a), minimal or no entanglement generation among Dirac particles is present, resulting in the absence of GC and CF. Notably, the trade-off between FOC and CF is evident here, without satisfying the upper bound relationship $D^2(\rho_{xyz})+F(\rho_{xyz})<1$.

When $T_{H}=1$, $T_{H}=10$, and $T_{H}=100$, as depicted in Fig. \ref{fig5}(b), (c), and (d) respectively, both CF and GC start appearing and reach the peak at $\alpha>1/\sqrt{2}$, and the minimum value of FOC decreases for $\alpha>1/\sqrt{2}$. From $T_{H}=10$  to $T_{H}=100$, one can find that the peak values of QC, GC, and CF do not increase significantly further.

Figure \ref{fig6}(a) illustrates the behaviors of QC, GC, CF, and FOC versus $T_H$ at $\omega=1$ and $\alpha=1/\sqrt{2}$. It is observed that with an increase in the Hawking temperature, QC, GC, and CF increase from zero to certain maximum values, thereafter stabilizing and persisting without reaching their respective maximum values. In contrast, FOC starts from its maximum, reaching a certain nonzero minimum value, and then saturating, while the trade-off relation between FOC and CF remains intact. Interestingly, this observation suggests that increasing Hawking temperature generates entanglement, in contrast to the completely accessible scenario where Hawking temperature degrades entanglement.

Unlike the effect of Hawking temperature on the aforementioned metrics, Fig. \ref{fig6}(b) showcases the behaviors of QC, GC, CF, and FOC concerning $\omega$ at $T_H=0.1$ and $\alpha=1/\sqrt{2}$. It is discerned that with an increase in mode frequency, QC, GC, and CF descend from their maximum values. Conversely, FOC increases from its minimum at higher mode frequencies,  all the while adhering to the trade-off between CF and FOC.

Let's analyze the mutual information shared between Alice--anti-Bob and Alice--anti-Charlie. In Figure \ref{fig7}(a), the variation of $I_{Ab}$ and $I_{Ac}$ as a function of $\alpha$ when $T_H=0.01$ and $\omega=1$ is depicted. We observe that both $I_{Ab}$ and $I_{Ac}$ are zero for all values of $\alpha$, which makes sense because when $T_H=0.01$, there is no generation of antiparticles inside the BH.

Now, when $T_H=10$, Fig. \ref{fig7}(b) demonstrates that both $I_{Ab}$ and $I_{Ac}$ become nonzero in general, indicating the creation of anti-Bob and anti-Charlie. The peak value of $I_{Ab}$ and $I_{Ac}$ appears to be around $\alpha\approx0.65$. It is notable that $\rho_{Ab}=\rho_{Ac}$, hence the mutual information for both pairs, $I_{Ab}$ and $I_{Ac}$, are equal.

Furthermore, in Fig. \ref{fig7}(c), we find that an increase in Hawking temperature from zero to $10$ generates $I_{Ab}=I_{Ac}$, and mutual information increases as temperature rises. Conversely, Fig. \ref{fig7}(d) shows the inverse relation of $I_{Ab}$ and $I_{Ac}$ with mode frequency so that the mutual information saturates to a zero value and does not change by increasing $\omega$.}

\begin{figure*}[!t]
\centering  \includegraphics[width=0.38\textwidth]{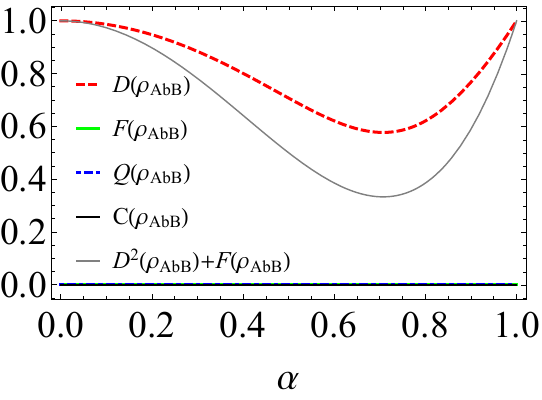}\put(-50,120){(a)}\qquad  \includegraphics[width=0.38\textwidth]{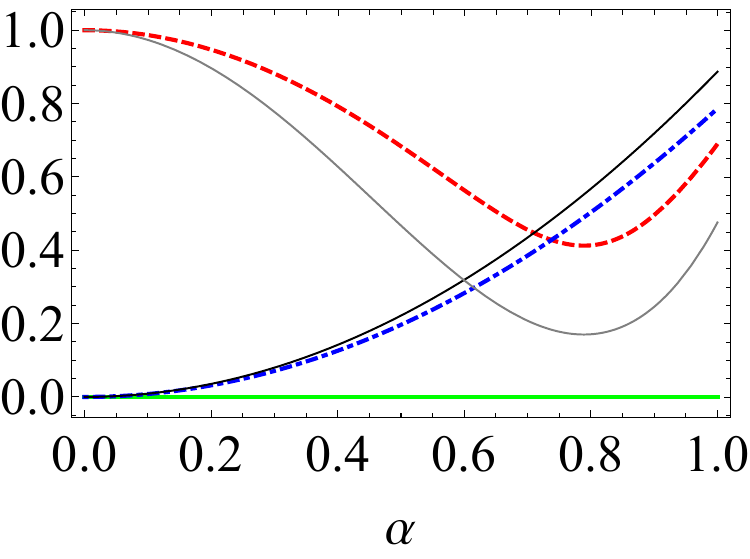} \put(-50,120){(b)}\qquad  \includegraphics[width=0.38\textwidth]{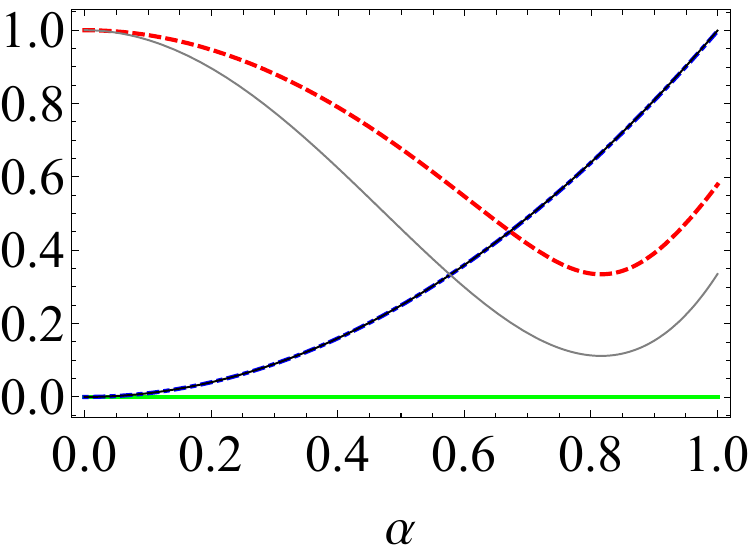}\put(-50,120){(c)}\qquad  \includegraphics[width=0.38\textwidth]{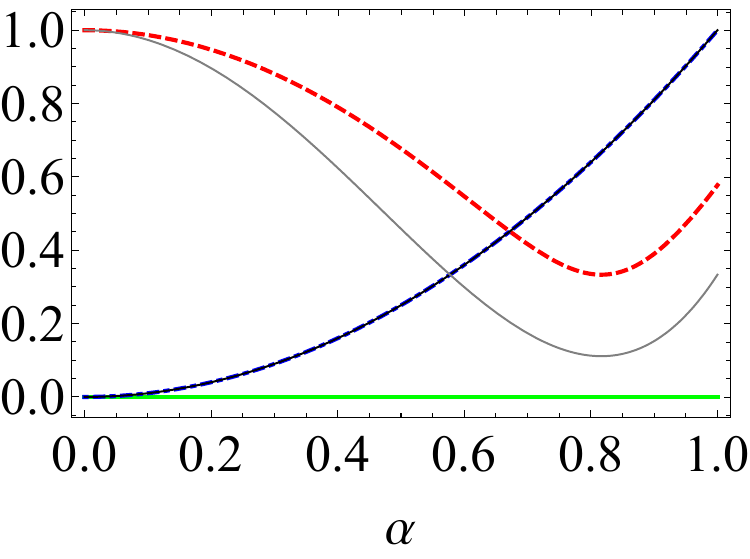}\put(-50,120){(d)}
\caption{FOC [$D(\rho_{AbB})$] (dashed-red), CF [$F(\rho_{AbB})$] (solid-green), GC [$Q(\rho_{AbB})$] (dot-dashed blue), QC [$C(\rho_{AbB})$] (solid-black) and $D^2(\rho_{AbB})+F(\rho_{AbB})$ (thin-solid gray)  as a function of $\alpha$ for $\omega=1$ at  $T_H=0.01$ (a), $T_H=1$ (b), $T_H=10$ (c),  and $T_H=100$ (d).}
\label{fig8}
\end{figure*}
\begin{figure*}[!t]
\begin{center}		\includegraphics[width=0.38\textwidth]{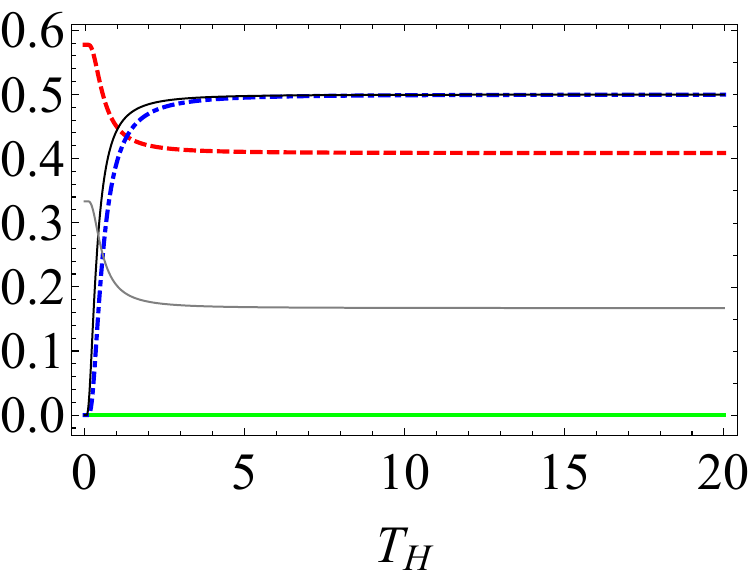}
		\put(-160,130){(a)}\qquad			\includegraphics[width=0.38\textwidth]{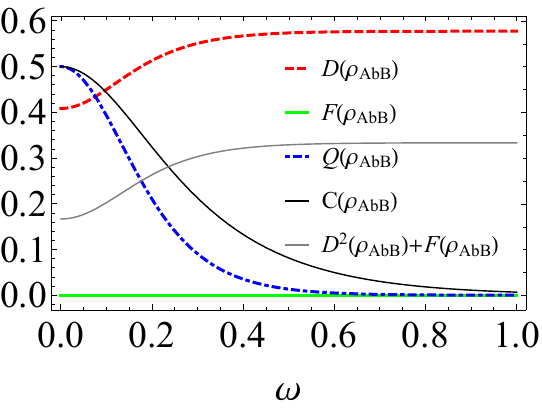}
		\put(-160,130){(b)}
	\end{center}
  \caption{FOC [$D(\rho_{AbB})$] (dashed-red), CF [$F(\rho_{AbB})$] (solid-green), GC [$Q(\rho_{AbB})$] (dot-dashed blue), QC [$C(\rho_{AbB})$] (Solid-black) and $D^2(\rho_{AbB})+F(\rho_{AbB})$ (thin-solid gray) as functions of $T_H$ and $\omega$ with $\alpha=1/\sqrt{2}$. (a) $\omega=1$   and (b)  $T_H=0.1$.}
  \label{fig9}
\end{figure*}

\begin{figure*}[!t]
\centering  \includegraphics[width=0.38\textwidth]{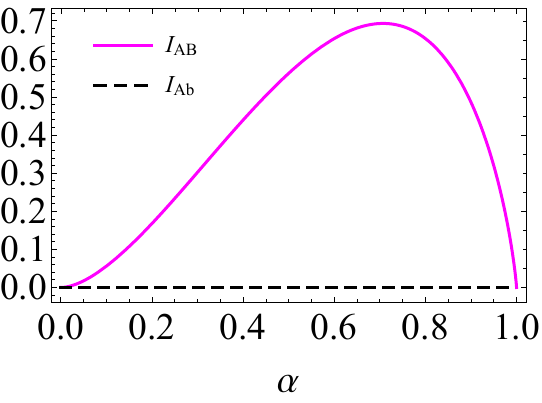}\put(-30,50){(a)}\qquad  \includegraphics[width=0.38\textwidth]{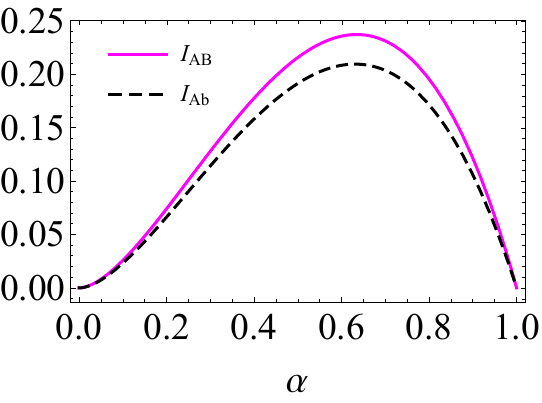} \put(-30,50){(b)}\qquad  \includegraphics[width=0.38\textwidth]{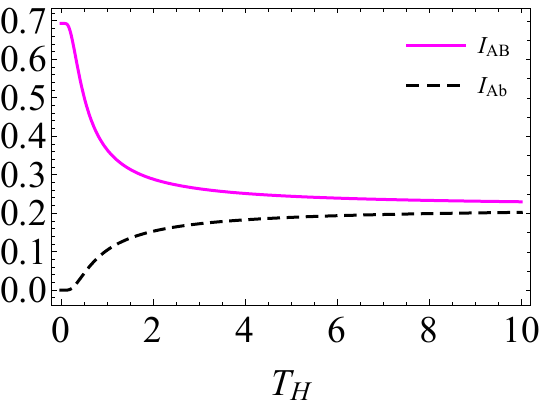}\put(-30,50){(c)}\qquad  \includegraphics[width=0.38\textwidth]{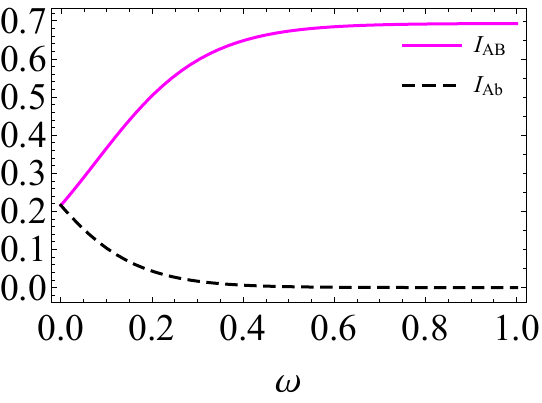}\put(-30,50){(d)}
\caption{Mutual information $I_{AB}$ and $I_{Ab}$ based on state \eqref{eq26}  with (a) $\omega=1$; $T_H=0.01$,  (b) $\omega=1$; $T_H=10$, (c) $\omega=1$; $\alpha=1/\sqrt{2}$,  and (d) $T_H=0.1$; $\alpha=1/\sqrt{2}$.}
\label{fig10}
\end{figure*}
\subsection{Alice--Bob--anti-Bob }\label{subsubsec:B}

As a third scenario, let us consider the partially accessible mode case comprised of Alice, anti-Bob, and Bob, whose density operator $\rho_{AbB}=\textmd{tr}_{cC}(\left|\psi\right\rangle_{AbBcC}\langle\psi|)$ would be represented as

\begin{align}
    \rho_{AbB}=&(\Theta_-^2 +\Gamma^2) \left|000\right\rangle \left\langle 000\right|+(\Theta_+^2 +\Gamma^2)\left|011\right\rangle \left\langle 011\right| \nonumber \\
    &+(\Theta_+ \Gamma +\Theta_- \Gamma)\{\left|000\right\rangle \left\langle 011\right| + \left|011\right\rangle \left\langle 000\right|\}\nonumber \\
    &+\Upsilon^2 \left|101\right\rangle \left\langle 101\right|.
     \label{eq26}
\end{align}

Unlike the accessible and completely inaccessible cases discussed in the previous subsections, QC, GC, CF, and FOC do not show the same trend in this partially accessible scenario.

{
In Fig. \ref{fig8}, FOC, CF, GC, and QC are depicted against $\alpha$ for various fixed values of Hawking temperature.
In Fig. \ref{fig8}(a), with $T_{H}=0.01$, negligible Hawking temperature results in no Dirac particle-antiparticle pair production, leading to minimal entanglement among Dirac particles and zero values for both CF and GC. However, FOC varies with $\alpha$, and the trade-off is evident.
Increasing the Hawking temperature to $T_{H}=1$ in Fig. \ref{fig8}(b) generates non-zero QC and GC, but CF remains at zero, indicating that the state is not genuinely entangled. A similar trend is observed in Fig. \ref{fig8}(c-d), but CF is consistently zero, meaning that in the case of Alice-Bob anti-Bob, no CF is created, resulting in no GME. However, there is global entanglement or quantum coherence despite this, and the trade-off relation strictly holds.

In Fig. \ref{fig9}(a), the variations of these measures with Hawking temperature at $\omega=1$ and $\alpha=1/\sqrt{2}$ are depicted. It is observed that CF remains consistently at zero, while QC and GC increase with rising Hawking temperature until reaching certain positive values where they saturate. Conversely, FOC exhibits the opposite behavior, decreasing with increasing Hawking temperature.
Figure \ref{fig9}(b) illustrates how all the measures change with mode frequency. It demonstrates that higher mode frequencies lead to a decrease in QC and GC while increasing FOC. Both GC and QC approach zero as the mode frequency tends to 1, maintaining a tight trade-off relationship for the given parameter values.

\begin{figure*}[t]
  \centering  \includegraphics[width=0.38\textwidth]{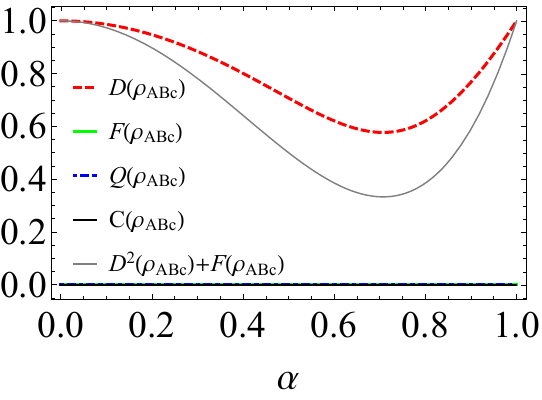}\put(-30,50){(a)}\qquad  \includegraphics[width=0.38\textwidth]{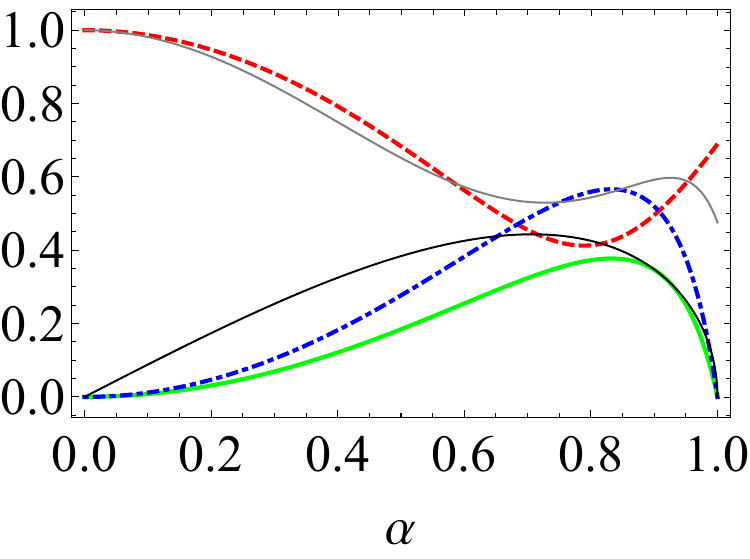} \put(-30,50){(b)}\qquad  \includegraphics[width=0.38\textwidth]{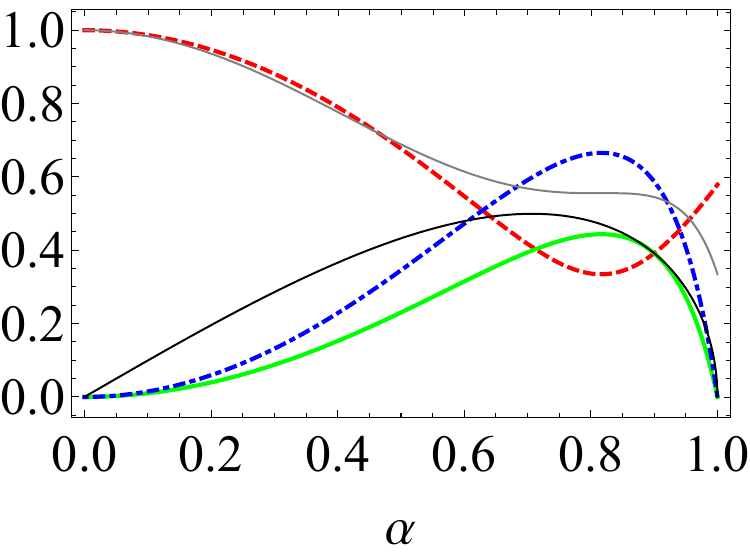}\put(-30,50){(c)}\qquad  \includegraphics[width=0.38\textwidth]{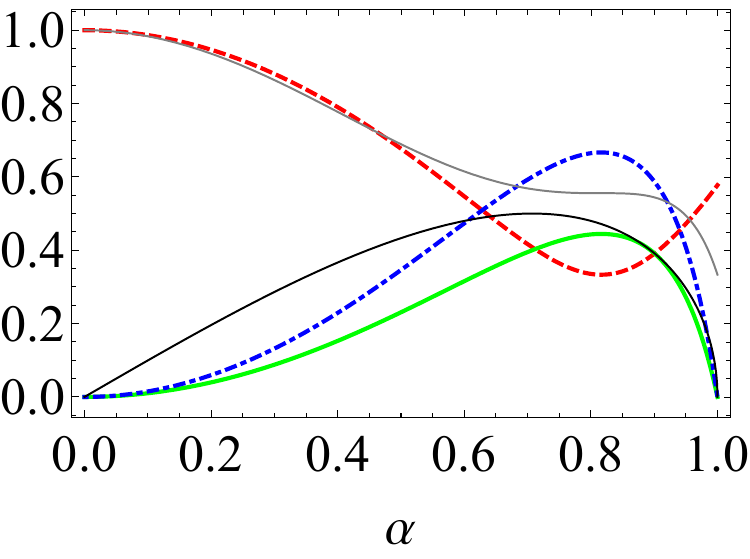}\put(-30,50){(d)}
\caption{FOC [$D(\rho_{ABc})$] (dashed-red), CF [$F(\rho_{ABc})$] (solid-green), GC [$Q(\rho_{ABc})$] (dot-dashed blue), QC [$C(\rho_{ABc})$] (solid-black) and $D^2(\rho_{ABc})+F(\rho_{ABc})$ (thin-solid gray)  as a function of $\alpha$ for $\omega=1$ at  $T_H=0.01$ (a), $T_H=1$ (b), $T_H=10$ (c),  and $T_H=100$ (d).}
\label{fig11}
\end{figure*}
\begin{figure*}[t]
\begin{center}		\includegraphics[width=0.38\textwidth]{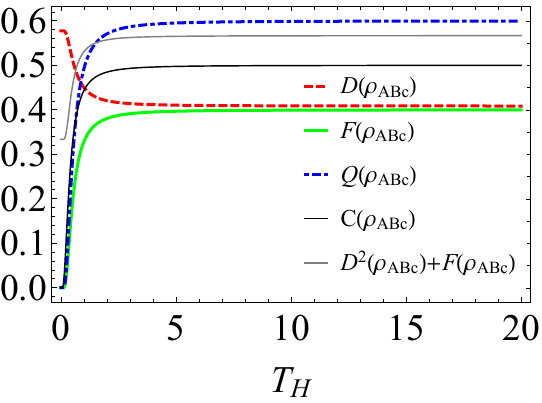}
\put(-160,50){(a)}\qquad	
\includegraphics[width=0.38\textwidth]{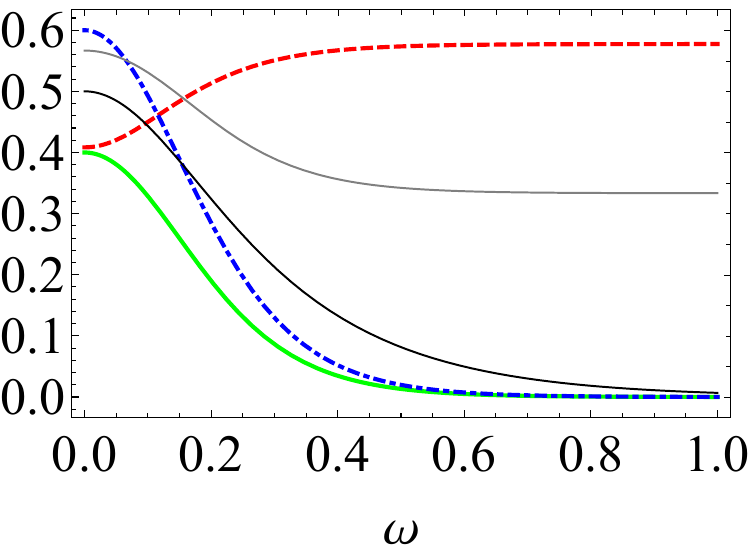}
\put(-160,50){(b)}
\end{center}
\caption{FOC [$D(\rho_{ABc})$] (dashed-red), CF [$F(\rho_{ABc})$] (solid-green), GC [$Q(\rho_{ABc})$] (dot-dashed blue), QC [$C(\rho_{ABc})$] (Solid-black) and $D^2(\rho_{ABc})+F(\rho_{ABc})$ (thin-solid gray) as functions of $T_H$ and $\omega$ with $\alpha=1/\sqrt{2}$. (a) $\omega=1$   and (b)  $T_H=0.1$.}
\label{fig12}
\end{figure*}
\begin{figure*}[!t]
\centering  \includegraphics[width=0.38\textwidth]{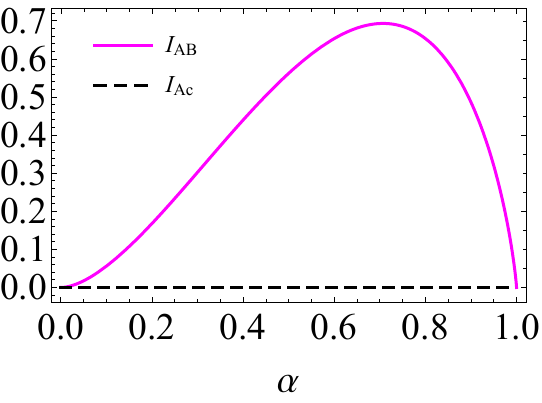}\put(-30,50){(a)}\qquad  \includegraphics[width=0.38\textwidth]{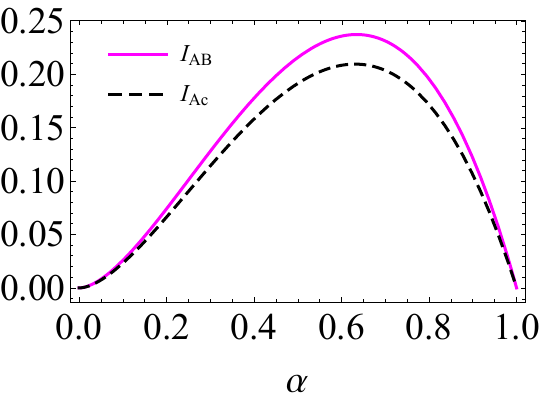} \put(-30,50){(b)}\qquad  \includegraphics[width=0.38\textwidth]{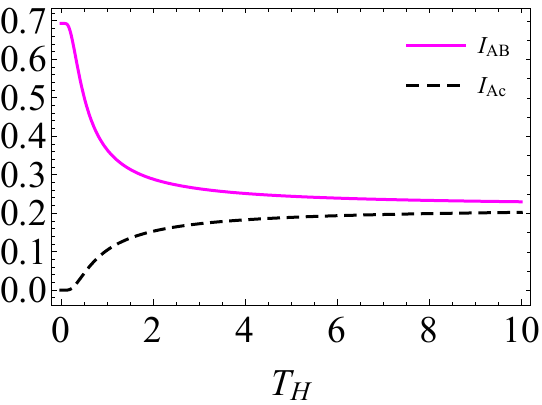}\put(-30,50){(c)}\qquad  \includegraphics[width=0.38\textwidth]{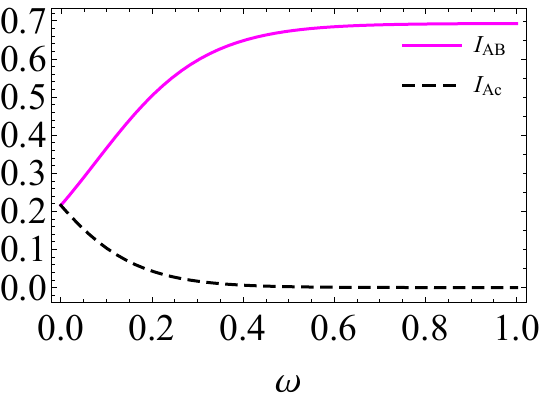}\put(-30,50){(d)}
\caption{Mutual information $I_{AB}$ and $I_{Ac}$ based on state \eqref{eq27} with (a) $\omega=1$; $T_H=0.01$,  (b) $\omega=1$; $T_H=10$, (c) $\omega=1$; $\alpha=1/\sqrt{2}$,  and (d) $T_H=0.1$; $\alpha=1/\sqrt{2}$.}
\label{fig13}
\end{figure*}

Let's examine the information correlation shared between $I_{AB}$ and $I_{Ab}$. In Fig. \ref{fig10}(a), the variations of $I_{AB}$ and $I_{Ab}$ as a function of $\alpha$ at $T_H=0.01$ and $\omega=1$ are illustrated. We observe that $I_{Ab}=0$ throughout for all values of $\alpha$, which makes sense as at very low Hawking temperatures no particles are generated inside the BH. However, the situation changes when $T_H=10$; both $I_{AB}$ and $I_{Ab}$ become nonzero, but still $I_{AB}>I_{Ab}$ except at $\alpha=0$ and $\alpha=1$.
In Fig. \ref{fig10}(c), the behaviors of $I_{AB}$ and $I_{Ab}$ as a function of $T_H$ at $\alpha=1/\sqrt{2}$ and $\omega=1$ are plotted. It shows that with an increase in Hawking temperature, $I_{AB}$ monotonically decreases, whereas $I_{Ab}$ increases from zero. For sufficiently large Hawking temperatures, both $I_{AB}$ and $I_{Ab}$ saturate to certain values with $I_{AB}>I_{Ab}$. Conversely, the divergence behavior of $I_{AB}$ and $I_{Ab}$ is seen when they are plotted against $\omega$, as shown in Fig. \ref{fig10}(d) at $T_H=0.1$ and $\alpha=1/\sqrt{2}$.

}

{ \subsection{Alice--Bob--anti-Charlie}}
{ Finally, as the fourth scenario, we consider an interesting case, Alice--Bob--anti-Charlie, with the following state
\begin{align}
\begin{split}
    \rho_{ABc} = & \Theta_-^2 \left|000\right\rangle \left\langle 000\right|+\Theta_+^2 \left|011\right\rangle \left\langle 011\right|  \\
    & + \Upsilon \Gamma \{\left|001\right\rangle \left\langle 110\right| +  \left|110\right\rangle \left\langle 001\right|\} \\
    & + \Gamma^2 \{\left|001\right\rangle \left\langle 001\right| +  \left|010\right\rangle \left\langle 010\right|\}\\
    & + \Upsilon^2 \left|110\right\rangle \left\langle 110\right|. \label{eq27}
\end{split}
\end{align}
}
{
Figure \ref{fig11} shows the variation of QC, GC, CF and FOC as a function of $\alpha$ with $\omega=1$.  At $T_H=0.01$, Fig. \ref{fig11}(a) shows that QC, GC and CF are all zero whereas FOC is equal to one at $\alpha=0$ and $\alpha=1$, and the trade-off relation holds. With the increase in Hawking temperature, i.e. $T_H=1$, we find that all the measures are now in general nonzero, meaning that the Hawking temperature is capable of generating QC, GC and CF. A similar trend is seen in Fig. \ref{fig11}(c-d).
Besides, Fig. \ref{fig12}(a) shows the behaviors of the mentioned measures as a function of Hawking temperature at $\alpha=1/\sqrt{2}$ and $\omega=1$, confirming that increasing the Hawking temperature increases QC, GC and CF from zero to some certain saturated values whereas FOC decreases from its maximum value. The opposite behavior is seen when these measures are plotted again mode frequency $\omega$, as shown in Fig. \ref{fig12}(b).

Let's examine the information correlation shared between $I_{AB}$ and $I_{Ac}$. In Fig. \ref{fig13}(a), the variations of $I_{AB}$ and $I_{Ac}$ as a function of $\alpha$ at $T_H=0.01$ and $\omega=1$ are illustrated. One can observe that $I_{Ac}=0$ throughout for all values of $\alpha$, which makes sense as for very low Hawking temperatures no particles are generated inside the BH. However, the situation changes when $T_H=10$ [see Fig. \ref{fig13}(b)], namely both $I_{AB}$ and $I_{Ac}$ become nonzero, but still $I_{AB}>I_{Ac}$ except for $\alpha=0$ and $\alpha=1$.
In Fig. \ref{fig13}(c), $I_{AB}$ and $I_{Ac}$ have been plotted as a function of $T_H$ at $\alpha=1/\sqrt{2}$ and $\omega=1$. It shows that with an increase in Hawking temperature, $I_{AB}$ monotonically decreases, whereas $I_{Ac}$ increases from zero.
For sufficiently large Hawking temperatures (when the BH approximates to evaporate completely, $T_H \to \infty$), the mutual information is distributed to the physically inaccessible region.
The converse behaviors of $I_{AB}$ and $I_{Ac}$ can be observed when they are plotted against $\omega$, as shown in Fig. \ref{fig13}(d).
}

\section{Summary and outlook}\label{sec5}
{ In this study, we conducted a comprehensive investigation of quantumness near a Schwarzschild black hole, examining various quantum resources and their interplay in curved space-time.  Our findings, depicted through multiple plots, reveal intriguing behaviors of quantum coherence, entanglement, and information correlation across different scenarios. In the accessible regime characterized by negligible Hawking temperature ($T_H=0.01$), the absence of particle-antiparticle pair production near the event horizon resulted in minimal entanglement among Dirac particles, reflected in zero values for quantum coherence, global concurrence, and concurrence fill. However, a persistent trade-off relationship between first-order coherence and concurrence fill underscored the intricate balance between coherence and entanglement. Transitioning to partially accessible scenarios with increasing Hawking temperature ($T_H=1$ to $T_H=100$), we observed the emergence of non-zero concurrence fill and global concurrence, indicating particle-antiparticle pair creation inside the black hole. Despite this, concurrence fill remained zero in certain scenarios, suggesting the absence of genuine entanglement. Notably, first-order coherence decreased with increasing Hawking temperature, while quantum coherence, concurrence fill and global concurrence exhibited saturation behavior, highlighting the coherence-entanglement trade-off. In the completely inaccessible scenario within the black hole's event horizon, the mutual information between external observers and particles inside the black hole became non-zero, signaling the creation of particle-antiparticle pairs. These findings deepen our understanding of quantum effects in curved space-time, shedding light on the quantum nature of black holes and paving the way for future investigations into the fundamental principles of quantum gravity.}

\appendix
\section{States of subsystems}
The states of all subsystems with $\Theta_{\pm} = \alpha S_{\pm}^2,$ $\Gamma =\alpha/2 \sqrt{\cosh ^2\left(\omega/2 T_H\right)}$, and $ \Upsilon = \sqrt{1-\alpha ^2}$ would be expressed as
\begin{equation}
    \rho_A=\left(
\begin{array}{cc}
 \Theta_+ ^2+2 \Gamma ^2+\Theta_- ^2 & 0 \\
 0 & \Upsilon^2 \\
\end{array}
\right),
\end{equation}
\begin{equation}
    \rho_B=\left(
\begin{array}{cc}
 \Gamma ^2+\Theta_- ^2 & 0 \\
 0 & \Theta_+ ^2+\Upsilon^2+\Gamma ^2 \\
\end{array}
\right),
\end{equation}
\begin{equation}
    \rho_b=\left(
\begin{array}{cc}
 \Upsilon^2+\Gamma ^2+\Theta_- ^2 & 0 \\
 0 & \Theta_+ ^2+\Gamma ^2 \\
\end{array}
\right),
\end{equation}
\begin{equation}
    \rho_C=\left(
\begin{array}{cc}
 \Gamma ^2+\Theta_- ^2 & 0 \\
 0 & \Theta_+ ^2+\Upsilon^2+\Gamma ^2 \\
\end{array}
\right),
\end{equation}
and
\begin{equation}
    \rho_c=\left(
\begin{array}{cc}
 \Upsilon^2+\Gamma ^2+\Theta_- ^2 & 0 \\
 0 & \Theta_+ ^2+\Gamma ^2 \\
\end{array}
\right).
\end{equation}
The state shared by Alice and Bob:
\begin{equation}
    \rho_{AB}=\left(
\begin{array}{cccc}
 \Gamma ^2+\Theta_- ^2 & 0 & 0 & 0 \\
 0 & \Theta_+ ^2+\Gamma ^2 & 0 & 0 \\
 0 & 0 & 0 & 0 \\
 0 & 0 & 0 & \Upsilon^2 \\
\end{array}
\right).
\end{equation}
The state shared by Alice and Charlie:
\begin{equation}
    \rho_{AC}=\left(
\begin{array}{cccc}
 \Gamma ^2+\Theta_- ^2 & 0 & 0 & 0 \\
 0 & \Theta_+ ^2+\Gamma ^2 & 0 & 0 \\
 0 & 0 & 0 & 0 \\
 0 & 0 & 0 & \Upsilon^2 \\
\end{array}
\right).
\end{equation}
The state shared by Bob and Charlie:
\begin{equation}
    \rho_{BC}=\left(
\begin{array}{cccc}
 \Theta_- ^2 & 0 & 0 & 0 \\
 0 & \Gamma ^2 & 0 & 0 \\
 0 & 0 & \Gamma ^2 & 0 \\
 0 & 0 & 0 & \Theta_+ ^2+\Upsilon^2 \\
\end{array}
\right).
\end{equation}
The state shared between Alice and anti-Bob:
\begin{equation}
    \rho_{Ab}=\left(
\begin{array}{cccc}
 \Gamma ^2+\Theta_- ^2 & 0 & 0 & 0 \\
 0 & \Theta_+ ^2+\Gamma ^2 & 0 & 0 \\
 0 & 0 & \Upsilon^2 & 0 \\
 0 & 0 & 0 & 0 \\
\end{array}
\right).
\end{equation}
The state shared between Bob and anti-Bob:
\begin{equation}
    \rho_{Bb}=\left(
\begin{array}{cccc}
 \Gamma ^2+\Theta_- ^2 & 0 & 0 & \Gamma  (\Theta_+ +\Theta_- ) \\
 0 & 0 & 0 & 0 \\
 0 & 0 & \Upsilon^2 & 0 \\
 \Gamma  (\Theta_+ +\Theta_- ) & 0 & 0 & \Theta_+ ^2+\Gamma ^2 \\
\end{array}
\right).
\end{equation}
The state shared between Alice and anti-Charlie:
\begin{equation}
    \rho_{Ac}=\left(
\begin{array}{cccc}
 \Gamma ^2+\Theta_- ^2 & 0 & 0 & 0 \\
 0 & \Theta_+ ^2+\Gamma ^2 & 0 & 0 \\
 0 & 0 & \Upsilon^2 & 0 \\
 0 & 0 & 0 & 0 \\
\end{array}
\right).
\end{equation}
The state shared between Bob and anti-Charlie:
\begin{equation}
    \rho_{Bc}=\left(
\begin{array}{cccc}
 \Theta_- ^2 & 0 & 0 & 0 \\
 0 & \Gamma ^2 & 0 & 0 \\
 0 & 0 & \Upsilon^2+\Gamma ^2 & 0 \\
 0 & 0 & 0 & \Theta_+ ^2 \\
\end{array}
\right).
\end{equation}
The state shared between Charlie and anti-Charlie:
\begin{equation}
\rho_{\text{cC}} = \begin{pmatrix}
    \Gamma^2 + \Theta_- ^2 & 0 & 0 & \Theta_+ \Gamma + \Theta_- \Gamma \\
    0 & \Upsilon^2 & 0 & 0 \\
    0 & 0 & 0 & 0 \\
    \Theta_+ \Gamma + \Theta_- \Gamma & 0 & 0 & \Theta_+ ^2 + \Gamma^2
\end{pmatrix}.
\end{equation}

\vspace{0.5cm}

\section*{Acknowledgements}
M.G. and S.H. were supported by Semnan University under Contract No. 21270. D. Wang was supported by the National Natural Science Foundation of China (Grant No. 12075001), and Anhui Provincial Key Research and Development Plan (Grant No. 2022b13020004).

\section*{Disclosures}
The authors declare that they have no known competing financial interests.

\section*{Data availability}
No datasets were generated or analyzed during the current study.

\bibliography{bibliography}

\begin{thebibliography}{44}%
\makeatletter
\providecommand \@ifxundefined [1]{%
 \@ifx{#1\undefined}
}%
\providecommand \@ifnum [1]{%
 \ifnum #1\expandafter \@firstoftwo
 \else \expandafter \@secondoftwo
 \fi
}%
\providecommand \@ifx [1]{%
 \ifx #1\expandafter \@firstoftwo
 \else \expandafter \@secondoftwo
 \fi
}%
\providecommand \natexlab [1]{#1}%
\providecommand \enquote  [1]{``#1''}%
\providecommand \bibnamefont  [1]{#1}%
\providecommand \bibfnamefont [1]{#1}%
\providecommand \citenamefont [1]{#1}%
\providecommand \href@noop [0]{\@secondoftwo}%
\providecommand \href [0]{\begingroup \@sanitize@url \@href}%
\providecommand \@href[1]{\@@startlink{#1}\@@href}%
\providecommand \@@href[1]{\endgroup#1\@@endlink}%
\providecommand \@sanitize@url [0]{\catcode `\\12\catcode `\$12\catcode
  `\&12\catcode `\#12\catcode `\^12\catcode `\_12\catcode `\%12\relax}%
\providecommand \@@startlink[1]{}%
\providecommand \@@endlink[0]{}%
\providecommand \url  [0]{\begingroup\@sanitize@url \@url }%
\providecommand \@url [1]{\endgroup\@href {#1}{\urlprefix }}%
\providecommand \urlprefix  [0]{URL }%
\providecommand \Eprint [0]{\href }%
\providecommand \doibase [0]{https://doi.org/}%
\providecommand \selectlanguage [0]{\@gobble}%
\providecommand \bibinfo  [0]{\@secondoftwo}%
\providecommand \bibfield  [0]{\@secondoftwo}%
\providecommand \translation [1]{[#1]}%
\providecommand \BibitemOpen [0]{}%
\providecommand \bibitemStop [0]{}%
\providecommand \bibitemNoStop [0]{.\EOS\space}%
\providecommand \EOS [0]{\spacefactor3000\relax}%
\providecommand \BibitemShut  [1]{\csname bibitem#1\endcsname}%
\let\auto@bib@innerbib\@empty
\bibitem [{\citenamefont
  {Schwarzschild}(1916)}]{schwarzschild1916gravitationsfeld}%
  \BibitemOpen
  \bibfield  {author} {\bibinfo {author} {\bibfnamefont {K.}~\bibnamefont
  {Schwarzschild}},\ }\bibfield  {title} {\bibinfo {title} {{\"U}ber das
  gravitationsfeld eines massenpunktes nach der einsteinschen theorie},\
  }\href@noop {} {\bibfield  {journal} {\bibinfo  {journal} {Sitzungsberichte
  der k{\"o}niglich preussischen Akademie der Wissenschaften}\ ,\ \bibinfo
  {pages} {189}} (\bibinfo {year} {1916})}\BibitemShut {NoStop}%
\bibitem [{\citenamefont {Akiyama}\ \emph {et~al.}(2019)\citenamefont
  {Akiyama}, \citenamefont {Alberdi}, \citenamefont {Alef}, \citenamefont
  {Asada}, \citenamefont {Azulay}, \citenamefont {Baczko}, \citenamefont
  {Ball}, \citenamefont {Balokovi{\'c}}, \citenamefont {Barrett}, \citenamefont
  {Bintley} \emph {et~al.}}]{akiyama2019first}%
  \BibitemOpen
  \bibfield  {author} {\bibinfo {author} {\bibfnamefont {K.}~\bibnamefont
  {Akiyama}}, \bibinfo {author} {\bibfnamefont {A.}~\bibnamefont {Alberdi}},
  \bibinfo {author} {\bibfnamefont {W.}~\bibnamefont {Alef}}, \bibinfo {author}
  {\bibfnamefont {K.}~\bibnamefont {Asada}}, \bibinfo {author} {\bibfnamefont
  {R.}~\bibnamefont {Azulay}}, \bibinfo {author} {\bibfnamefont {A.-K.}\
  \bibnamefont {Baczko}}, \bibinfo {author} {\bibfnamefont {D.}~\bibnamefont
  {Ball}}, \bibinfo {author} {\bibfnamefont {M.}~\bibnamefont {Balokovi{\'c}}},
  \bibinfo {author} {\bibfnamefont {J.}~\bibnamefont {Barrett}}, \bibinfo
  {author} {\bibfnamefont {D.}~\bibnamefont {Bintley}}, \emph {et~al.},\
  }\bibfield  {title} {\bibinfo {title} {First {M}87 event horizon telescope
  results. iv. imaging the central supermassive black hole},\ }\href@noop {}
  {\bibfield  {journal} {\bibinfo  {journal} {The Astrophysical Journal
  Letters}\ }\textbf {\bibinfo {volume} {875}},\ \bibinfo {pages} {L4}
  (\bibinfo {year} {2019})}\BibitemShut {NoStop}%
\bibitem [{\citenamefont {G{\"u}rlebeck}(2015)}]{gurlebeck2015no}%
  \BibitemOpen
  \bibfield  {author} {\bibinfo {author} {\bibfnamefont {N.}~\bibnamefont
  {G{\"u}rlebeck}},\ }\bibfield  {title} {\bibinfo {title} {No-hair theorem for
  black holes in astrophysical environments},\ }\href@noop {} {\bibfield
  {journal} {\bibinfo  {journal} {Physical Review Letters}\ }\textbf {\bibinfo
  {volume} {114}},\ \bibinfo {pages} {151102} (\bibinfo {year}
  {2015})}\BibitemShut {NoStop}%
\bibitem [{\citenamefont {Hawking}(1974)}]{hawking1974black}%
  \BibitemOpen
  \bibfield  {author} {\bibinfo {author} {\bibfnamefont {S.~W.}\ \bibnamefont
  {Hawking}},\ }\bibfield  {title} {\bibinfo {title} {Black hole explosions?},\
  }\href@noop {} {\bibfield  {journal} {\bibinfo  {journal} {Nature}\ }\textbf
  {\bibinfo {volume} {248}},\ \bibinfo {pages} {30} (\bibinfo {year}
  {1974})}\BibitemShut {NoStop}%
\bibitem [{\citenamefont {Denis}(2023)}]{denis2023entropy}%
  \BibitemOpen
  \bibfield  {author} {\bibinfo {author} {\bibfnamefont {O.}~\bibnamefont
  {Denis}},\ }\bibfield  {title} {\bibinfo {title} {The entropy of the
  entangled {H}awking radiation},\ }\href@noop {} {\bibfield  {journal}
  {\bibinfo  {journal} {IPI Letters (1)}\ ,\ \bibinfo {pages} {1}} (\bibinfo
  {year} {2023})}\BibitemShut {NoStop}%
\bibitem [{\citenamefont {Almheiri}\ \emph {et~al.}(2021)\citenamefont
  {Almheiri}, \citenamefont {Hartman}, \citenamefont {Maldacena}, \citenamefont
  {Shaghoulian},\ and\ \citenamefont {Tajdini}}]{almheiri2021entropy}%
  \BibitemOpen
  \bibfield  {author} {\bibinfo {author} {\bibfnamefont {A.}~\bibnamefont
  {Almheiri}}, \bibinfo {author} {\bibfnamefont {T.}~\bibnamefont {Hartman}},
  \bibinfo {author} {\bibfnamefont {J.}~\bibnamefont {Maldacena}}, \bibinfo
  {author} {\bibfnamefont {E.}~\bibnamefont {Shaghoulian}},\ and\ \bibinfo
  {author} {\bibfnamefont {A.}~\bibnamefont {Tajdini}},\ }\bibfield  {title}
  {\bibinfo {title} {The entropy of {H}awking radiation},\ }\href@noop {}
  {\bibfield  {journal} {\bibinfo  {journal} {Reviews of Modern Physics}\
  }\textbf {\bibinfo {volume} {93}},\ \bibinfo {pages} {035002} (\bibinfo
  {year} {2021})}\BibitemShut {NoStop}%
\bibitem [{\citenamefont {Iizuka}\ and\ \citenamefont
  {Kabat}(2013)}]{PhysRevD.88.084010}%
  \BibitemOpen
  \bibfield  {author} {\bibinfo {author} {\bibfnamefont {N.}~\bibnamefont
  {Iizuka}}\ and\ \bibinfo {author} {\bibfnamefont {D.}~\bibnamefont {Kabat}},\
  }\bibfield  {title} {\bibinfo {title} {Mutual information in {H}awking
  radiation},\ }\href@noop {} {\bibfield  {journal} {\bibinfo  {journal} {Phys.
  Rev. D}\ }\textbf {\bibinfo {volume} {88}},\ \bibinfo {pages} {084010}
  (\bibinfo {year} {2013})}\BibitemShut {NoStop}%
\bibitem [{\citenamefont {Wang}\ \emph {et~al.}(2018)\citenamefont {Wang},
  \citenamefont {Shi}, \citenamefont {Hoehn}, \citenamefont {Ming},
  \citenamefont {Sun}, \citenamefont {Kais},\ and\ \citenamefont
  {Ye}}]{DongWangADP2018}%
  \BibitemOpen
  \bibfield  {author} {\bibinfo {author} {\bibfnamefont {D.}~\bibnamefont
  {Wang}}, \bibinfo {author} {\bibfnamefont {W.-N.}\ \bibnamefont {Shi}},
  \bibinfo {author} {\bibfnamefont {R.~D.}\ \bibnamefont {Hoehn}}, \bibinfo
  {author} {\bibfnamefont {F.}~\bibnamefont {Ming}}, \bibinfo {author}
  {\bibfnamefont {W.-Y.}\ \bibnamefont {Sun}}, \bibinfo {author} {\bibfnamefont
  {S.}~\bibnamefont {Kais}},\ and\ \bibinfo {author} {\bibfnamefont
  {L.}~\bibnamefont {Ye}},\ }\bibfield  {title} {\bibinfo {title} {Effects of
  {H}awking radiation on the entropic uncertainty in a {S}chwarzschild
  space-time},\ }\href@noop {} {\bibfield  {journal} {\bibinfo  {journal}
  {Annalen der Physik}\ }\textbf {\bibinfo {volume} {530}},\ \bibinfo {pages}
  {1800080} (\bibinfo {year} {2018})}\BibitemShut {NoStop}%
\bibitem [{\citenamefont {Huang}\ \emph {et~al.}(2018)\citenamefont {Huang},
  \citenamefont {Ma}, \citenamefont {Wang},\ and\ \citenamefont
  {Ye}}]{Huang2018Hawking}%
  \BibitemOpen
  \bibfield  {author} {\bibinfo {author} {\bibfnamefont {C.~Y.}\ \bibnamefont
  {Huang}}, \bibinfo {author} {\bibfnamefont {W.~C.}\ \bibnamefont {Ma}},
  \bibinfo {author} {\bibfnamefont {D.}~\bibnamefont {Wang}},\ and\ \bibinfo
  {author} {\bibfnamefont {L.}~\bibnamefont {Ye}},\ }\bibfield  {title}
  {\bibinfo {title} {How the {H}awking radiation affect quantum {F}isher
  information of {D}irac particles in the background of a schwarzschild black
  hole},\ }\href@noop {} {\bibfield  {journal} {\bibinfo  {journal} {Quantum
  Inf. Process.}\ }\textbf {\bibinfo {volume} {17}},\ \bibinfo {pages} {16}
  (\bibinfo {year} {2018})}\BibitemShut {NoStop}%
\bibitem [{\citenamefont {Shi}\ \emph {et~al.}(2018)\citenamefont {Shi},
  \citenamefont {Ding}, \citenamefont {He}, \citenamefont {Yu}, \citenamefont
  {Wu}, \citenamefont {Chen}, \citenamefont {Wang}, \citenamefont {Liu},
  \citenamefont {Sun},\ and\ \citenamefont {Ye}}]{SHI2018649}%
  \BibitemOpen
  \bibfield  {author} {\bibinfo {author} {\bibfnamefont {J.}~\bibnamefont
  {Shi}}, \bibinfo {author} {\bibfnamefont {Z.}~\bibnamefont {Ding}}, \bibinfo
  {author} {\bibfnamefont {J.}~\bibnamefont {He}}, \bibinfo {author}
  {\bibfnamefont {L.}~\bibnamefont {Yu}}, \bibinfo {author} {\bibfnamefont
  {T.}~\bibnamefont {Wu}}, \bibinfo {author} {\bibfnamefont {S.}~\bibnamefont
  {Chen}}, \bibinfo {author} {\bibfnamefont {D.}~\bibnamefont {Wang}}, \bibinfo
  {author} {\bibfnamefont {C.}~\bibnamefont {Liu}}, \bibinfo {author}
  {\bibfnamefont {W.}~\bibnamefont {Sun}},\ and\ \bibinfo {author}
  {\bibfnamefont {L.}~\bibnamefont {Ye}},\ }\bibfield  {title} {\bibinfo
  {title} {Quantum distinguishability and geometric discord in the background
  of {S}chwarzschild space–time},\ }\href@noop {} {\bibfield  {journal}
  {\bibinfo  {journal} {Physica A: Statistical Mechanics and its Applications}\
  }\textbf {\bibinfo {volume} {510}},\ \bibinfo {pages} {649} (\bibinfo {year}
  {2018})}\BibitemShut {NoStop}%
\bibitem [{\citenamefont {Li}\ \emph {et~al.}(2022)\citenamefont {Li},
  \citenamefont {Ming}, \citenamefont {Song}, \citenamefont {Ye},\ and\
  \citenamefont {Wang}}]{li2022quantumness}%
  \BibitemOpen
  \bibfield  {author} {\bibinfo {author} {\bibfnamefont {L.-J.}\ \bibnamefont
  {Li}}, \bibinfo {author} {\bibfnamefont {F.}~\bibnamefont {Ming}}, \bibinfo
  {author} {\bibfnamefont {X.-K.}\ \bibnamefont {Song}}, \bibinfo {author}
  {\bibfnamefont {L.}~\bibnamefont {Ye}},\ and\ \bibinfo {author}
  {\bibfnamefont {D.}~\bibnamefont {Wang}},\ }\bibfield  {title} {\bibinfo
  {title} {Quantumness and entropic uncertainty in curved space-time},\
  }\href@noop {} {\bibfield  {journal} {\bibinfo  {journal} {The European
  Physical Journal C}\ }\textbf {\bibinfo {volume} {82}},\ \bibinfo {pages}
  {726} (\bibinfo {year} {2022})}\BibitemShut {NoStop}%
\bibitem [{\citenamefont {Wu}\ and\ \citenamefont
  {Zeng}(2022{\natexlab{a}})}]{Wu2022EPJC}%
  \BibitemOpen
  \bibfield  {author} {\bibinfo {author} {\bibfnamefont {S.~M.}\ \bibnamefont
  {Wu}}\ and\ \bibinfo {author} {\bibfnamefont {H.~S.}\ \bibnamefont {Zeng}},\
  }\bibfield  {title} {\bibinfo {title} {Genuine tripartite nonlocality and
  entanglement in curved spacetime},\ }\href@noop {} {\bibfield  {journal}
  {\bibinfo  {journal} {Eur. Phys. J. C}\ }\textbf {\bibinfo {volume} {82}},\
  \bibinfo {pages} {4} (\bibinfo {year} {2022}{\natexlab{a}})}\BibitemShut
  {NoStop}%
\bibitem [{\citenamefont {Wu}\ and\ \citenamefont
  {Zeng}(2022{\natexlab{b}})}]{Wu2022EPJC2}%
  \BibitemOpen
  \bibfield  {author} {\bibinfo {author} {\bibfnamefont {S.~M.}\ \bibnamefont
  {Wu}}\ and\ \bibinfo {author} {\bibfnamefont {H.~S.}\ \bibnamefont {Zeng}},\
  }\bibfield  {title} {\bibinfo {title} {Fermionic steering and its monogamy
  relations in {S}chwarzschild spacetime},\ }\href@noop {} {\bibfield
  {journal} {\bibinfo  {journal} {Eur. Phys. J. C}\ }\textbf {\bibinfo {volume}
  {82}},\ \bibinfo {pages} {716} (\bibinfo {year}
  {2022}{\natexlab{b}})}\BibitemShut {NoStop}%
\bibitem [{\citenamefont {Chen}\ \emph {et~al.}(2023)\citenamefont {Chen},
  \citenamefont {Hu},\ and\ \citenamefont {Yu}}]{wu2023monogamy}%
  \BibitemOpen
  \bibfield  {author} {\bibinfo {author} {\bibfnamefont {Y.}~\bibnamefont
  {Chen}}, \bibinfo {author} {\bibfnamefont {J.}~\bibnamefont {Hu}},\ and\
  \bibinfo {author} {\bibfnamefont {H.}~\bibnamefont {Yu}},\ }\bibfield
  {title} {\bibinfo {title} {Collective transitions of two entangled atoms near
  a {S}chwarzschild black hole},\ }\href@noop {} {\bibfield  {journal}
  {\bibinfo  {journal} {Physical Review D}\ }\textbf {\bibinfo {volume}
  {107}},\ \bibinfo {pages} {025015} (\bibinfo {year} {2023})}\BibitemShut
  {NoStop}%
\bibitem [{\citenamefont {Wang}(2023)}]{wang2023nonperturbative}%
  \BibitemOpen
  \bibfield  {author} {\bibinfo {author} {\bibfnamefont {L.}~\bibnamefont
  {Wang}},\ }\bibfield  {title} {\bibinfo {title} {A nonperturbative approach
  to {H}awking radiation and black hole quantum hair},\ }\href@noop {}
  {\bibfield  {journal} {\bibinfo  {journal} {Classical and Quantum Gravity}\
  }\textbf {\bibinfo {volume} {40}},\ \bibinfo {pages} {225010} (\bibinfo
  {year} {2023})}\BibitemShut {NoStop}%
\bibitem [{\citenamefont {Fujita}\ and\ \citenamefont
  {Zhang}(2023)}]{fujita2023holographic}%
  \BibitemOpen
  \bibfield  {author} {\bibinfo {author} {\bibfnamefont {M.}~\bibnamefont
  {Fujita}}\ and\ \bibinfo {author} {\bibfnamefont {J.}~\bibnamefont {Zhang}},\
  }\bibfield  {title} {\bibinfo {title} {Holographic entanglement entropy of
  the double {W}ick rotated {BTZ} black hole},\ }\href@noop {} {\bibfield
  {journal} {\bibinfo  {journal} {Physical Review D}\ }\textbf {\bibinfo
  {volume} {107}},\ \bibinfo {pages} {026007} (\bibinfo {year}
  {2023})}\BibitemShut {NoStop}%
\bibitem [{\citenamefont {Wu}\ \emph {et~al.}(2023{\natexlab{a}})\citenamefont
  {Wu}, \citenamefont {Fan}, \citenamefont {Huang},\ and\ \citenamefont
  {Zeng}}]{Wu_2023}%
  \BibitemOpen
  \bibfield  {author} {\bibinfo {author} {\bibfnamefont {S.-M.}\ \bibnamefont
  {Wu}}, \bibinfo {author} {\bibfnamefont {X.-W.}\ \bibnamefont {Fan}},
  \bibinfo {author} {\bibfnamefont {X.-L.}\ \bibnamefont {Huang}},\ and\
  \bibinfo {author} {\bibfnamefont {H.-S.}\ \bibnamefont {Zeng}},\ }\bibfield
  {title} {\bibinfo {title} {Genuine tripartite entanglement of {W} state
  subject to {H}awking effect of a {S}chwarzschild black hole},\ }\href@noop {}
  {\bibfield  {journal} {\bibinfo  {journal} {Europhysics Letters}\ }\textbf
  {\bibinfo {volume} {141}},\ \bibinfo {pages} {18001} (\bibinfo {year}
  {2023}{\natexlab{a}})}\BibitemShut {NoStop}%
\bibitem [{\citenamefont {Zhang}\ \emph {et~al.}(2023)\citenamefont {Zhang},
  \citenamefont {Wang},\ and\ \citenamefont {Fei}}]{EPJC2023Fei}%
  \BibitemOpen
  \bibfield  {author} {\bibinfo {author} {\bibfnamefont {T.}~\bibnamefont
  {Zhang}}, \bibinfo {author} {\bibfnamefont {X.}~\bibnamefont {Wang}},\ and\
  \bibinfo {author} {\bibfnamefont {S.-M.}\ \bibnamefont {Fei}},\ }\bibfield
  {title} {\bibinfo {title} {Hawking effect can generate physically
  inaccessible genuine tripartite nonlocality},\ }\href@noop {} {\bibfield
  {journal} {\bibinfo  {journal} {Eur. Phys. J. C}\ }\textbf {\bibinfo {volume}
  {83}},\ \bibinfo {pages} {607} (\bibinfo {year} {2023})}\BibitemShut
  {NoStop}%
\bibitem [{\citenamefont {Xu}\ \emph {et~al.}(2014)\citenamefont {Xu},
  \citenamefont {Song}, \citenamefont {Shi},\ and\ \citenamefont
  {Ye}}]{xu2014hawking}%
  \BibitemOpen
  \bibfield  {author} {\bibinfo {author} {\bibfnamefont {S.}~\bibnamefont
  {Xu}}, \bibinfo {author} {\bibfnamefont {X.-k.}\ \bibnamefont {Song}},
  \bibinfo {author} {\bibfnamefont {J.-d.}\ \bibnamefont {Shi}},\ and\ \bibinfo
  {author} {\bibfnamefont {L.}~\bibnamefont {Ye}},\ }\bibfield  {title}
  {\bibinfo {title} {How the {H}awking effect affects multipartite entanglement
  of {D}irac particles in the background of a {S}chwarzschild black hole},\
  }\href@noop {} {\bibfield  {journal} {\bibinfo  {journal} {Physical Review
  D}\ }\textbf {\bibinfo {volume} {89}},\ \bibinfo {pages} {065022} (\bibinfo
  {year} {2014})}\BibitemShut {NoStop}%
\bibitem [{\citenamefont {Mart{\'\i}n-Mart{\'\i}nez}\ \emph
  {et~al.}(2010)\citenamefont {Mart{\'\i}n-Mart{\'\i}nez}, \citenamefont
  {Garay},\ and\ \citenamefont {Le{\'o}n}}]{martin2010unveiling}%
  \BibitemOpen
  \bibfield  {author} {\bibinfo {author} {\bibfnamefont {E.}~\bibnamefont
  {Mart{\'\i}n-Mart{\'\i}nez}}, \bibinfo {author} {\bibfnamefont {L.~J.}\
  \bibnamefont {Garay}},\ and\ \bibinfo {author} {\bibfnamefont
  {J.}~\bibnamefont {Le{\'o}n}},\ }\bibfield  {title} {\bibinfo {title}
  {Unveiling quantum entanglement degradation near a {S}chwarzschild black
  hole},\ }\href@noop {} {\bibfield  {journal} {\bibinfo  {journal} {Physical
  review D}\ }\textbf {\bibinfo {volume} {82}},\ \bibinfo {pages} {064006}
  (\bibinfo {year} {2010})}\BibitemShut {NoStop}%
\bibitem [{\citenamefont {Wang}\ \emph
  {et~al.}(2010{\natexlab{a}})\citenamefont {Wang}, \citenamefont {Pan},\ and\
  \citenamefont {Jing}}]{wang2010entanglement}%
  \BibitemOpen
  \bibfield  {author} {\bibinfo {author} {\bibfnamefont {J.}~\bibnamefont
  {Wang}}, \bibinfo {author} {\bibfnamefont {Q.}~\bibnamefont {Pan}},\ and\
  \bibinfo {author} {\bibfnamefont {J.}~\bibnamefont {Jing}},\ }\bibfield
  {title} {\bibinfo {title} {Entanglement redistribution in the {S}chwarzschild
  spacetime},\ }\href@noop {} {\bibfield  {journal} {\bibinfo  {journal}
  {Physics Letters B}\ }\textbf {\bibinfo {volume} {692}},\ \bibinfo {pages}
  {202} (\bibinfo {year} {2010}{\natexlab{a}})}\BibitemShut {NoStop}%
\bibitem [{\citenamefont {Wang}\ \emph
  {et~al.}(2010{\natexlab{b}})\citenamefont {Wang}, \citenamefont {Pan},\ and\
  \citenamefont {Jing}}]{wang2010projective}%
  \BibitemOpen
  \bibfield  {author} {\bibinfo {author} {\bibfnamefont {J.}~\bibnamefont
  {Wang}}, \bibinfo {author} {\bibfnamefont {Q.}~\bibnamefont {Pan}},\ and\
  \bibinfo {author} {\bibfnamefont {J.}~\bibnamefont {Jing}},\ }\bibfield
  {title} {\bibinfo {title} {Projective measurements and generation of
  entangled {D}irac particles in {S}chwarzschild spacetime},\ }\href@noop {}
  {\bibfield  {journal} {\bibinfo  {journal} {Annals of Physics}\ }\textbf
  {\bibinfo {volume} {325}},\ \bibinfo {pages} {1190} (\bibinfo {year}
  {2010}{\natexlab{b}})}\BibitemShut {NoStop}%
\bibitem [{\citenamefont {Haddadi}\ \emph {et~al.}(2024)\citenamefont
  {Haddadi}, \citenamefont {Yurischev}, \citenamefont {Abd-Rabbou},
  \citenamefont {Azizi}, \citenamefont {Pourkarimi},\ and\ \citenamefont
  {Ghominejad}}]{haddadi2024}%
  \BibitemOpen
  \bibfield  {author} {\bibinfo {author} {\bibfnamefont {S.}~\bibnamefont
  {Haddadi}}, \bibinfo {author} {\bibfnamefont {M.~A.}\ \bibnamefont
  {Yurischev}}, \bibinfo {author} {\bibfnamefont {M.~Y.}\ \bibnamefont
  {Abd-Rabbou}}, \bibinfo {author} {\bibfnamefont {M.}~\bibnamefont {Azizi}},
  \bibinfo {author} {\bibfnamefont {M.~R.}\ \bibnamefont {Pourkarimi}},\ and\
  \bibinfo {author} {\bibfnamefont {M.}~\bibnamefont {Ghominejad}},\ }\bibfield
   {title} {\bibinfo {title} {Quantumness near a {S}chwarzschild black hole},\
  }\href@noop {} {\bibfield  {journal} {\bibinfo  {journal} {Eur. Phys. J. C}\
  }\textbf {\bibinfo {volume} {84}},\ \bibinfo {pages} {42} (\bibinfo {year}
  {2024})}\BibitemShut {NoStop}%
\bibitem [{\citenamefont {Hawking}(2005)}]{hawking2005information}%
  \BibitemOpen
  \bibfield  {author} {\bibinfo {author} {\bibfnamefont {S.~W.}\ \bibnamefont
  {Hawking}},\ }\bibfield  {title} {\bibinfo {title} {Information loss in black
  holes},\ }\href@noop {} {\bibfield  {journal} {\bibinfo  {journal} {Physical
  Review D}\ }\textbf {\bibinfo {volume} {72}},\ \bibinfo {pages} {084013}
  (\bibinfo {year} {2005})}\BibitemShut {NoStop}%
\bibitem [{\citenamefont {Xie}\ and\ \citenamefont
  {Eberly}(2021)}]{xie2021triangle}%
  \BibitemOpen
  \bibfield  {author} {\bibinfo {author} {\bibfnamefont {S.}~\bibnamefont
  {Xie}}\ and\ \bibinfo {author} {\bibfnamefont {J.~H.}\ \bibnamefont
  {Eberly}},\ }\bibfield  {title} {\bibinfo {title} {Triangle measure of
  tripartite entanglement},\ }\href@noop {} {\bibfield  {journal} {\bibinfo
  {journal} {Physical Review Letters}\ }\textbf {\bibinfo {volume} {127}},\
  \bibinfo {pages} {040403} (\bibinfo {year} {2021})}\BibitemShut {NoStop}%
\bibitem [{\citenamefont {Meyer}\ and\ \citenamefont
  {Wallach}(2002)}]{meyer2002global}%
  \BibitemOpen
  \bibfield  {author} {\bibinfo {author} {\bibfnamefont {D.~A.}\ \bibnamefont
  {Meyer}}\ and\ \bibinfo {author} {\bibfnamefont {N.~R.}\ \bibnamefont
  {Wallach}},\ }\bibfield  {title} {\bibinfo {title} {Global entanglement in
  multiparticle systems},\ }\href@noop {} {\bibfield  {journal} {\bibinfo
  {journal} {Journal of Mathematical Physics}\ }\textbf {\bibinfo {volume}
  {43}},\ \bibinfo {pages} {4273} (\bibinfo {year} {2002})}\BibitemShut
  {NoStop}%
\bibitem [{\citenamefont {Brennen}(2003)}]{brennen2003observable}%
  \BibitemOpen
  \bibfield  {author} {\bibinfo {author} {\bibfnamefont {G.~K.}\ \bibnamefont
  {Brennen}},\ }\bibfield  {title} {\bibinfo {title} {An observable measure of
  entanglement for pure states of multi-qubit systems},\ }\href@noop {}
  {\bibfield  {journal} {\bibinfo  {journal} {arXiv preprint quant-ph/0305094}\
  } (\bibinfo {year} {2003})}\BibitemShut {NoStop}%
\bibitem [{\citenamefont {Svozil\'{\i}k}\ \emph {et~al.}(2015)\citenamefont
  {Svozil\'{\i}k}, \citenamefont {Vall\'es}, \citenamefont
  {Pe\ifmmode~\check{r}\else \v{r}\fi{}ina},\ and\ \citenamefont
  {Torres}}]{PhysRevLett.115.220501}%
  \BibitemOpen
  \bibfield  {author} {\bibinfo {author} {\bibfnamefont {J.}~\bibnamefont
  {Svozil\'{\i}k}}, \bibinfo {author} {\bibfnamefont {A.}~\bibnamefont
  {Vall\'es}}, \bibinfo {author} {\bibfnamefont {J.}~\bibnamefont
  {Pe\ifmmode~\check{r}\else \v{r}\fi{}ina}},\ and\ \bibinfo {author}
  {\bibfnamefont {J.~P.}\ \bibnamefont {Torres}},\ }\bibfield  {title}
  {\bibinfo {title} {Revealing hidden coherence in partially coherent light},\
  }\href@noop {} {\bibfield  {journal} {\bibinfo  {journal} {Phys. Rev. Lett.}\
  }\textbf {\bibinfo {volume} {115}},\ \bibinfo {pages} {220501} (\bibinfo
  {year} {2015})}\BibitemShut {NoStop}%
\bibitem [{\citenamefont {Wu}\ \emph {et~al.}(2023{\natexlab{b}})\citenamefont
  {Wu}, \citenamefont {Liu}, \citenamefont {Wang}, \citenamefont {Li},
  \citenamefont {Huang},\ and\ \citenamefont {Zeng}}]{svozilik2015revealing}%
  \BibitemOpen
  \bibfield  {author} {\bibinfo {author} {\bibfnamefont {S.-M.}\ \bibnamefont
  {Wu}}, \bibinfo {author} {\bibfnamefont {D.-D.}\ \bibnamefont {Liu}},
  \bibinfo {author} {\bibfnamefont {C.-X.}\ \bibnamefont {Wang}}, \bibinfo
  {author} {\bibfnamefont {W.-M.}\ \bibnamefont {Li}}, \bibinfo {author}
  {\bibfnamefont {X.-L.}\ \bibnamefont {Huang}},\ and\ \bibinfo {author}
  {\bibfnamefont {H.-S.}\ \bibnamefont {Zeng}},\ }\bibfield  {title} {\bibinfo
  {title} {Monogamy relationship between quantum and classical correlations for
  continuous variable in curved spacetime},\ }\href@noop {} {\bibfield
  {journal} {\bibinfo  {journal} {The European Physical Journal Plus}\ }\textbf
  {\bibinfo {volume} {138}},\ \bibinfo {pages} {56} (\bibinfo {year}
  {2023}{\natexlab{b}})}\BibitemShut {NoStop}%
\bibitem [{\citenamefont {Ali}\ \emph {et~al.}(2021)\citenamefont {Ali},
  \citenamefont {Nadeem},\ and\ \citenamefont {Toor}}]{ali2021properties}%
  \BibitemOpen
  \bibfield  {author} {\bibinfo {author} {\bibfnamefont {A.}~\bibnamefont
  {Ali}}, \bibinfo {author} {\bibfnamefont {M.}~\bibnamefont {Nadeem}},\ and\
  \bibinfo {author} {\bibfnamefont {A.}~\bibnamefont {Toor}},\ }\bibfield
  {title} {\bibinfo {title} {Properties of quantum coherence and correlations
  in quasi-entangled coherent states},\ }\href@noop {} {\bibfield  {journal}
  {\bibinfo  {journal} {The European Physical Journal D}\ }\textbf {\bibinfo
  {volume} {75}},\ \bibinfo {pages} {266} (\bibinfo {year} {2021})}\BibitemShut
  {NoStop}%
\bibitem [{\citenamefont {Dong}\ \emph {et~al.}(2022)\citenamefont {Dong},
  \citenamefont {Wei}, \citenamefont {Song}, \citenamefont {Wang},\ and\
  \citenamefont {Ye}}]{dong2022unification}%
  \BibitemOpen
  \bibfield  {author} {\bibinfo {author} {\bibfnamefont {D.-D.}\ \bibnamefont
  {Dong}}, \bibinfo {author} {\bibfnamefont {G.-B.}\ \bibnamefont {Wei}},
  \bibinfo {author} {\bibfnamefont {X.-K.}\ \bibnamefont {Song}}, \bibinfo
  {author} {\bibfnamefont {D.}~\bibnamefont {Wang}},\ and\ \bibinfo {author}
  {\bibfnamefont {L.}~\bibnamefont {Ye}},\ }\bibfield  {title} {\bibinfo
  {title} {Unification of coherence and quantum correlations in tripartite
  systems},\ }\href@noop {} {\bibfield  {journal} {\bibinfo  {journal}
  {Physical Review A}\ }\textbf {\bibinfo {volume} {106}},\ \bibinfo {pages}
  {042415} (\bibinfo {year} {2022})}\BibitemShut {NoStop}%
\bibitem [{\citenamefont {Du}\ and\ \citenamefont
  {Tong}(2021)}]{PhysRevA.103.032407}%
  \BibitemOpen
  \bibfield  {author} {\bibinfo {author} {\bibfnamefont {M.-M.}\ \bibnamefont
  {Du}}\ and\ \bibinfo {author} {\bibfnamefont {D.~M.}\ \bibnamefont {Tong}},\
  }\bibfield  {title} {\bibinfo {title} {Relationship between first-order
  coherence and the maximum violation of the three-setting linear steering
  inequality for a two-qubit system},\ }\href@noop {} {\bibfield  {journal}
  {\bibinfo  {journal} {Phys. Rev. A}\ }\textbf {\bibinfo {volume} {103}},\
  \bibinfo {pages} {032407} (\bibinfo {year} {2021})}\BibitemShut {NoStop}%
\bibitem [{\citenamefont {Guo}(2024)}]{guo2023complete}%
  \BibitemOpen
  \bibfield  {author} {\bibinfo {author} {\bibfnamefont {Y.}~\bibnamefont
  {Guo}},\ }\bibfield  {title} {\bibinfo {title} {Complete genuine multipartite
  entanglement monotone},\ }\href@noop {} {\bibfield  {journal} {\bibinfo
  {journal} {Results in Physics}\ }\textbf {\bibinfo {volume} {57}},\ \bibinfo
  {pages} {107430} (\bibinfo {year} {2024})}\BibitemShut {NoStop}%
\bibitem [{\citenamefont {Jin}\ \emph {et~al.}(2023)\citenamefont {Jin},
  \citenamefont {Tao}, \citenamefont {Gui}, \citenamefont {Fei}, \citenamefont
  {Li-Jost},\ and\ \citenamefont {Qiao}}]{jin2023concurrence}%
  \BibitemOpen
  \bibfield  {author} {\bibinfo {author} {\bibfnamefont {Z.-X.}\ \bibnamefont
  {Jin}}, \bibinfo {author} {\bibfnamefont {Y.-H.}\ \bibnamefont {Tao}},
  \bibinfo {author} {\bibfnamefont {Y.-T.}\ \bibnamefont {Gui}}, \bibinfo
  {author} {\bibfnamefont {S.-M.}\ \bibnamefont {Fei}}, \bibinfo {author}
  {\bibfnamefont {X.}~\bibnamefont {Li-Jost}},\ and\ \bibinfo {author}
  {\bibfnamefont {C.-F.}\ \bibnamefont {Qiao}},\ }\bibfield  {title} {\bibinfo
  {title} {Concurrence triangle induced genuine multipartite entanglement
  measure},\ }\href@noop {} {\bibfield  {journal} {\bibinfo  {journal} {Results
  in Physics}\ }\textbf {\bibinfo {volume} {44}},\ \bibinfo {pages} {106155}
  (\bibinfo {year} {2023})}\BibitemShut {NoStop}%
\bibitem [{\citenamefont {Baumgratz}\ \emph {et~al.}(2014)\citenamefont
  {Baumgratz}, \citenamefont {Cramer},\ and\ \citenamefont
  {Plenio}}]{baumgratz2014quantifying}%
  \BibitemOpen
  \bibfield  {author} {\bibinfo {author} {\bibfnamefont {T.}~\bibnamefont
  {Baumgratz}}, \bibinfo {author} {\bibfnamefont {M.}~\bibnamefont {Cramer}},\
  and\ \bibinfo {author} {\bibfnamefont {M.~B.}\ \bibnamefont {Plenio}},\
  }\bibfield  {title} {\bibinfo {title} {Quantifying coherence},\ }\href@noop
  {} {\bibfield  {journal} {\bibinfo  {journal} {Physical review letters}\
  }\textbf {\bibinfo {volume} {113}},\ \bibinfo {pages} {140401} (\bibinfo
  {year} {2014})}\BibitemShut {NoStop}%
\bibitem [{\citenamefont {Hu}\ \emph {et~al.}(2018)\citenamefont {Hu},
  \citenamefont {Hu}, \citenamefont {Wang}, \citenamefont {Peng}, \citenamefont
  {Zhang},\ and\ \citenamefont {Fan}}]{HU20181}%
  \BibitemOpen
  \bibfield  {author} {\bibinfo {author} {\bibfnamefont {M.-L.}\ \bibnamefont
  {Hu}}, \bibinfo {author} {\bibfnamefont {X.}~\bibnamefont {Hu}}, \bibinfo
  {author} {\bibfnamefont {J.}~\bibnamefont {Wang}}, \bibinfo {author}
  {\bibfnamefont {Y.}~\bibnamefont {Peng}}, \bibinfo {author} {\bibfnamefont
  {Y.-R.}\ \bibnamefont {Zhang}},\ and\ \bibinfo {author} {\bibfnamefont
  {H.}~\bibnamefont {Fan}},\ }\bibfield  {title} {\bibinfo {title} {Quantum
  coherence and geometric quantum discord},\ }\href@noop {} {\bibfield
  {journal} {\bibinfo  {journal} {Physics Reports}\ }\textbf {\bibinfo {volume}
  {762-764}},\ \bibinfo {pages} {1} (\bibinfo {year} {2018})}\BibitemShut
  {NoStop}%
\bibitem [{\citenamefont {Tessier}(2005)}]{tessier2005complementarity}%
  \BibitemOpen
  \bibfield  {author} {\bibinfo {author} {\bibfnamefont {T.~E.}\ \bibnamefont
  {Tessier}},\ }\bibfield  {title} {\bibinfo {title} {Complementarity relations
  for multi-qubit systems},\ }\href@noop {} {\bibfield  {journal} {\bibinfo
  {journal} {Foundations of Physics Letters}\ }\textbf {\bibinfo {volume}
  {18}},\ \bibinfo {pages} {107} (\bibinfo {year} {2005})}\BibitemShut
  {NoStop}%
\bibitem [{\citenamefont {Fan}\ \emph {et~al.}(2019)\citenamefont {Fan},
  \citenamefont {Sun}, \citenamefont {Ding}, \citenamefont {Ming},
  \citenamefont {Yang}, \citenamefont {Wang},\ and\ \citenamefont
  {Ye}}]{fan2019universal}%
  \BibitemOpen
  \bibfield  {author} {\bibinfo {author} {\bibfnamefont {X.-G.}\ \bibnamefont
  {Fan}}, \bibinfo {author} {\bibfnamefont {W.-Y.}\ \bibnamefont {Sun}},
  \bibinfo {author} {\bibfnamefont {Z.-Y.}\ \bibnamefont {Ding}}, \bibinfo
  {author} {\bibfnamefont {F.}~\bibnamefont {Ming}}, \bibinfo {author}
  {\bibfnamefont {H.}~\bibnamefont {Yang}}, \bibinfo {author} {\bibfnamefont
  {D.}~\bibnamefont {Wang}},\ and\ \bibinfo {author} {\bibfnamefont
  {L.}~\bibnamefont {Ye}},\ }\bibfield  {title} {\bibinfo {title} {Universal
  complementarity between coherence and intrinsic concurrence for two-qubit
  states},\ }\href@noop {} {\bibfield  {journal} {\bibinfo  {journal} {New
  Journal of Physics}\ }\textbf {\bibinfo {volume} {21}},\ \bibinfo {pages}
  {093053} (\bibinfo {year} {2019})}\BibitemShut {NoStop}%
\bibitem [{\citenamefont {Dong}\ \emph {et~al.}(2023)\citenamefont {Dong},
  \citenamefont {Song}, \citenamefont {Fan}, \citenamefont {Ye},\ and\
  \citenamefont {Wang}}]{PhysRevA.107.052403}%
  \BibitemOpen
  \bibfield  {author} {\bibinfo {author} {\bibfnamefont {D.-D.}\ \bibnamefont
  {Dong}}, \bibinfo {author} {\bibfnamefont {X.-K.}\ \bibnamefont {Song}},
  \bibinfo {author} {\bibfnamefont {X.-G.}\ \bibnamefont {Fan}}, \bibinfo
  {author} {\bibfnamefont {L.}~\bibnamefont {Ye}},\ and\ \bibinfo {author}
  {\bibfnamefont {D.}~\bibnamefont {Wang}},\ }\bibfield  {title} {\bibinfo
  {title} {Complementary relations of entanglement, coherence, steering, and
  {B}ell nonlocality inequality violation in three-qubit states},\ }\href@noop
  {} {\bibfield  {journal} {\bibinfo  {journal} {Phys. Rev. A}\ }\textbf
  {\bibinfo {volume} {107}},\ \bibinfo {pages} {052403} (\bibinfo {year}
  {2023})}\BibitemShut {NoStop}%
\bibitem [{\citenamefont {Guo}\ \emph {et~al.}(2022)\citenamefont {Guo},
  \citenamefont {Jia}, \citenamefont {Li},\ and\ \citenamefont
  {Huang}}]{guo2022genuine}%
  \BibitemOpen
  \bibfield  {author} {\bibinfo {author} {\bibfnamefont {Y.}~\bibnamefont
  {Guo}}, \bibinfo {author} {\bibfnamefont {Y.}~\bibnamefont {Jia}}, \bibinfo
  {author} {\bibfnamefont {X.}~\bibnamefont {Li}},\ and\ \bibinfo {author}
  {\bibfnamefont {L.}~\bibnamefont {Huang}},\ }\bibfield  {title} {\bibinfo
  {title} {Genuine multipartite entanglement measure},\ }\href@noop {}
  {\bibfield  {journal} {\bibinfo  {journal} {Journal of Physics A:
  Mathematical and Theoretical}\ }\textbf {\bibinfo {volume} {55}},\ \bibinfo
  {pages} {145303} (\bibinfo {year} {2022})}\BibitemShut {NoStop}%
\bibitem [{\citenamefont {Zhu}\ and\ \citenamefont
  {Fei}(2015)}]{zhu2015generalized}%
  \BibitemOpen
  \bibfield  {author} {\bibinfo {author} {\bibfnamefont {X.-N.}\ \bibnamefont
  {Zhu}}\ and\ \bibinfo {author} {\bibfnamefont {S.-M.}\ \bibnamefont {Fei}},\
  }\bibfield  {title} {\bibinfo {title} {Generalized monogamy relations of
  concurrence for {N}-qubit systems},\ }\href@noop {} {\bibfield  {journal}
  {\bibinfo  {journal} {Physical Review A}\ }\textbf {\bibinfo {volume} {92}},\
  \bibinfo {pages} {062345} (\bibinfo {year} {2015})}\BibitemShut {NoStop}%
\bibitem [{\citenamefont {Damour}\ and\ \citenamefont
  {Ruffini}(1976)}]{damour1976black}%
  \BibitemOpen
  \bibfield  {author} {\bibinfo {author} {\bibfnamefont {T.}~\bibnamefont
  {Damour}}\ and\ \bibinfo {author} {\bibfnamefont {R.}~\bibnamefont
  {Ruffini}},\ }\bibfield  {title} {\bibinfo {title} {Black-hole evaporation in
  the {K}lein-{S}auter-{H}eisenberg-{E}uler formalism},\ }\href@noop {}
  {\bibfield  {journal} {\bibinfo  {journal} {Physical Review D}\ }\textbf
  {\bibinfo {volume} {14}},\ \bibinfo {pages} {332} (\bibinfo {year}
  {1976})}\BibitemShut {NoStop}%
\bibitem [{\citenamefont {Bogoljubov}\ \emph {et~al.}(1958)\citenamefont
  {Bogoljubov}, \citenamefont {Tolmachov},\ and\ \citenamefont
  {{\v{S}}irkov}}]{bogoljubov1958new}%
  \BibitemOpen
  \bibfield  {author} {\bibinfo {author} {\bibfnamefont {N.}~\bibnamefont
  {Bogoljubov}}, \bibinfo {author} {\bibfnamefont {V.~V.}\ \bibnamefont
  {Tolmachov}},\ and\ \bibinfo {author} {\bibfnamefont {D.}~\bibnamefont
  {{\v{S}}irkov}},\ }\bibfield  {title} {\bibinfo {title} {A new method in the
  theory of superconductivity},\ }\href@noop {} {\bibfield  {journal} {\bibinfo
   {journal} {Fortschritte der Physik}\ }\textbf {\bibinfo {volume} {6}},\
  \bibinfo {pages} {605} (\bibinfo {year} {1958})}\BibitemShut {NoStop}%
\bibitem [{\citenamefont {Barnett}\ and\ \citenamefont
  {Radmore}(2002)}]{barnett2002methods}%
  \BibitemOpen
  \bibfield  {author} {\bibinfo {author} {\bibfnamefont {S.}~\bibnamefont
  {Barnett}}\ and\ \bibinfo {author} {\bibfnamefont {P.~M.}\ \bibnamefont
  {Radmore}},\ }\href@noop {} {\emph {\bibinfo {title} {Methods in theoretical
  quantum optics}}},\ Vol.~\bibinfo {volume} {15}\ (\bibinfo  {publisher}
  {Oxford University Press},\ \bibinfo {year} {2002})\BibitemShut {NoStop}%
\end{thebibliography}%

\end{document}